\documentclass{vki}

\usepackage{lmodern,pdftexcmds,amsmath}

\usepackage{amsfonts,amsthm,bm} % Math packages

\DeclareMathAlphabet{\mathsfbi}{OT1}{\sfdefault}{bx}{sl}
\DeclareMathVersion{sfletters}
\SetSymbolFont{letters}{sfletters}{OML}{ntxsfmi}{b}{it}

\makeatletter
\newcommand{\mathbfsbilow}[1]{%
	\text{\mathversion{sfletters}$\m@th#1$}%
}

\DeclareRobustCommand{\tensor}[1]{%
	\begingroup
	\ifcat\noexpand #1\relax
	\edef\greek@test{\detokenize{#1}}%
	\edef\greek@test{\expandafter\@cdr\greek@test\@nil}%
	\edef\greek@test{\expandafter\@car\greek@test\@nil}%
	\edef\x{\the\lccode\expandafter`\greek@test}%
	\edef\y{\number\expandafter`\greek@test}%
	\ifnum\x=\y\relax
	\mathbfsbilow{#1}%
	\else
	\mathsfbi{#1}%
	\fi
	\else
	\mathsfbi{#1}%
	\fi
	\endgroup
}
\makeatother

\usepackage[most]{tcolorbox}
\usepackage{booktabs}
\usepackage{amsmath}
\usepackage{wrapfig}
\usepackage[linesnumbered,ruled,vlined]{algorithm2e}% http://ctan.org/pkg/algorithm2e
\usepackage{algpseudocode}
\usepackage{pdflscape}
\usepackage{subcaption}

\usepackage{siunitx}
\usepackage{comment}

\usepackage{mathtools}
\tcbset{enhanced,colback=cyan!2!white,colframe=cyan!90!white,coltitle=blue!20!black,fonttitle=\bfseries}

\usepackage{listings}
\usepackage{hyperref}
\hypersetup{
	colorlinks=true,
	linkcolor=blue,
	filecolor=magenta,      
	citecolor=blue,
	urlcolor=cyan,
}

\begin{document}

% USER ENTRY ON
%\layout
% uncomment this \layout to have an idea of the margins
% USER ENTRY OFF

\pagenumbering{roman}
% USER ENTRY ON
\title{Learning with Physical Constraints (Chapter 3)}
\author{Miguel A. Mendez\thanks{mendez@vki.ac.be}, Jan van Den Berghe, Manuel Ratz, \\Matilde Fiore and Lorenzo Schena\\von Karman Institute for Fluid Dynamics}

\date{3 December 2024} % I suggest you adjust this manually
% USER ENTRY OFF
\maketitle

This chapter complements the previous by providing three tutorial exercises on physics-constrained regression. These are implemented as ``toy problems'' that seek to mimic grand challenges in (1) the super-resolution and data assimilation of the velocity field in image velocimetry, (2) data-driven turbulence modeling, and (3) system identification and digital twinning for forecasting and control. The Python codes for all exercises are provided in the course repository.

\vspace{50mm}

How to cite this work
\bigskip

\begin{centering}
	\begin{lstlisting}
@InCollection{Mendez2024,
  author    = {Mendez, Miguel A. and Van den Berghe, Jan and Ratz, Manuel
               and Fiore, Matilde and Schena, Lorenzo},
  title     = {Learning with Physical Constraints},
  booktitle = {Machine Learning for Fluid Dynamics},
  editor    = {Mendez, Miguel A. and Parente, Alessandro},
  publisher = {von Karman Institute},
  year      = {2024},
  chapter   = {3},
  isbn      = {978-2875162090}
}
	\end{lstlisting}
\end{centering}

%%%%%%%%%%%%%%%%%%%%%%
\pagenumbering{arabic}
\setcounter{page}{1}
\clearpage{\pagestyle{empty}} 

\tableofcontents
\clearpage{\pagestyle{empty}} 

\vspace{-3mm}

\section{Problem 1: "Fill the Gaps" and super-resolution } \label{ch_3_sec_4}

\subsection{General Context}\label{ch_3_sec_1_1}

In experimental fluid mechanics, velocity fields are often measured using image velocimetry. This includes traditional Particle Image Velocimetry (PIV, \cite{Raffel2018}), which relies on cross-correlation to estimate particle displacement on a regular grid, and Particle Tracking Velocimetry (PTV, \cite{Schroeder2023}), in which individual particles are localized and tracked to obtain velocity measurements at scattered spatial locations. Both approaches require post-processing to address missing or unreliable data: filtering and outlier removal in PIV can create gaps in the grid, while PTV data are inherently scattered. Historically, PIV gaps were filled using non-parametric local averaging methods (see Chapter 2). However, with the increasing adoption of three-dimensional velocimetry techniques \citep{Elsinga2006,Schanz2016}, the emphasis has shifted toward parametric and physics-informed regression methods for reconstructing complete flow fields.

The key advantage of a parametric model is its analytical representation, which allows for compact storage and facilitates the enforcement of constraints such as exact divergence-free conditions in incompressible flows. The resulting analytic velocity field can be evaluated at \emph{any} spatial location and provides symbolic (rather than numerical) derivatives, a capability often referred to as super-resolution. This removes the need to interpolate the data onto a specific grid or to compute derivatives using finite-difference approximations. The incorporation of physical priors or governing equations in the reconstruction process is commonly referred to as data assimilation, a field that has grown substantially in recent years (see \cite{Gesemann2016,Schneiders2016,Agarwal2021,Sperotto2022a,Sciacchitano2022b,Jeon2022}). The present "toy problem" therefore serves as an introductory exercise in data assimilation for image velocimetry.

\begin{figure}[ht]
\center
 \includegraphics[width=0.7\textwidth]{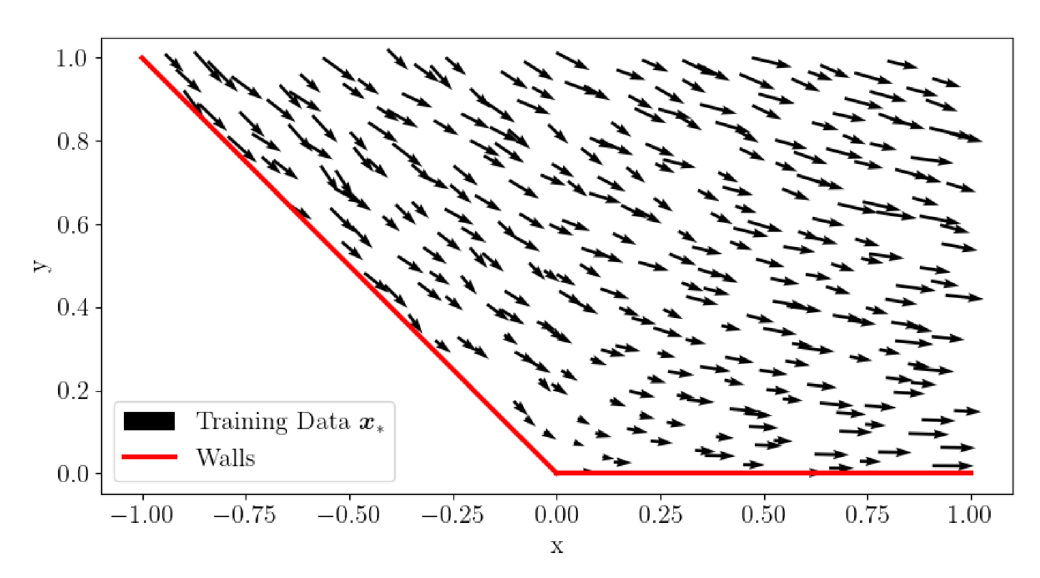}
    \centering
    \caption{Quiver plot of the training data in the constrained RBF regression exercise of Section \ref{ch_3_sec_4}.}
    \label{fig:ex1_training_data}
\end{figure}

\subsection{Proposed Exercise}\label{ch_3_sec_1_2}

In this exercise, we make use of the RBF-constrained framework proposed in \citet{Sperotto2022a} and implemented in the open-source software SPICY\footnote{\url{https://github.com/mendezVKI/SPICY_VKI}} (meshlesS Pressure from Image veloCimetrY), developed at VKI \citep{Sperotto2024}. The version provided to participants includes recent extensions contributed by Manuel Ratz, who currently leads ongoing developments on meshless data assimilation for turbulence statistics (see \citet{Ratz2024}).

In the terminology of Section 2 from chapter 2, the method belongs to the third category, in which \emph{hard constraints} are used to enforce physical priors during regression. The task is to reconstruct a noisy velocity field measured in a two-dimensional corner flow while imposing (1) boundary conditions, (2) divergence-free constraints, and (3) curl-free conditions. In addition to these hard constraints, we also include a global penalty term promoting divergence-free behavior. As we will briefly discuss, enforcing constraints increases memory requirements and numerical difficulty, whereas penalty terms introduce no additional memory cost and can therefore assist the regression.

The flow considered here is shown in Figure~\ref{fig:ex1_training_data}. Training samples are depicted as velocity arrows, and the walls are shown in red. The reference flow is a 2D potential corner flow with potential $A r^n \cos(n\Theta)$, where $A = 1$ and $n = 4/3$, corresponding to a corner angle of $135^\circ$. We sample the resulting velocity field at 298 quasi-random scattered points and add 30\% uniform, uncorrelated noise to emulate measurement conditions. While this level of noise is higher than what is typically encountered in practice, it serves to clearly illustrate the robustness of constrained regression.

\subsection{Methodology and Results}\label{ch_3_sec_4_3}

The constrained RBF approach consists of minimizing the augmented cost function 

\begin{equation}
\label{Augmented}
    \mathcal{A}(\bm{w},\bm{\lambda})= \mathcal{J}(\bm{w}) + \alpha \mathcal{P}(\bm{w}) + \bm{\lambda}^T \mathcal{H}(\bm{w})\,,
\end{equation} where $\mathcal{P}$ is a quadratic penalty, $\mathcal{H}$ a linear constraint, and $\bm{w}$ are the weights for the RBF regression (as in 13 from Chapter 2). The reason for taking the penalties as quadratic functions and the constraints as linear functions is that by doing so, setting the gradients $d\mathcal{A}/d\bm{w}=0$ and  $d\mathcal{A}/d\bm{\lambda}=0$ leads to a linear system. Moreover, the minimization must treat both $\bm{w}$ and $\bm{\lambda}$ as unknown. Let us analyze each of the terms in \eqref{Augmented} independently.

The first term is the data-driven cost function. We consider a standard $l_2$ norm at the scope. For the regression of a vector field with training data $\bm{w}_U^*=\bm{u}(\bm{x}^*)=(u(\bm{x}^*),v(\bm{x}^*))$ this reads:

\begin{equation}
    \mathcal{J}(\bm{w}_U) = \left\vert\left\vert\begin{pmatrix}
        \bm{\Phi}(\bm{x_{*}}) & \bm{0} \\
        \bm{0} & \bm{\Phi}(\bm{x_{*}})
    \end{pmatrix} \begin{pmatrix}
        \bm{w}_u \\
        \bm{w}_v
    \end{pmatrix} - \begin{pmatrix}
        \bm{u}_{*} \\
        \bm{v}_{*}
    \end{pmatrix}\right\vert\right\vert_2^2 = \vert\vert \bm{\Phi}_U(\bm{x}_{*}) \bm{w}_U - \bm{U}) \vert\vert_2^2\,,
\end{equation} where we have shortened the expression through the notation on the right-hand side.

As a quadratic penalty, we consider the divergence-free condition. This requires the spatial derivatives. For the RBF model, these are easily available in any point $\bm{x}$ by differentiating the basis functions: 
\begin{align}
    \partial_x u(\bm{x}) &= \sum^{n_b-1}_{r=0} w_r \partial_x \phi_r(\bm{x}) = \bm{\Phi}_x(\bm{x}) \bm{w}_u \\
    \partial_y v(\bm{x}) &= \sum^{n_b-1}_{r=0} w_r \partial_y\phi_r(\bm{x}) = \bm{\Phi}_y(\bm{x}) \bm{w}_u\,.
    \label{eq:rbf_cost}
\end{align}

Here, the columns of $\bm{\Phi}_x$ and $\bm{\Phi}_y$ collect the derivative of each RBF along the $x$ and $y$ direction, respectively. With these analytical derivatives, we can now define a quadratic penalty term as: 
\begin{equation}
    \mathcal{P} = \left\vert\left\vert\begin{pmatrix}
        \bm{\Phi}_x(\bm{x}_{*}) & \bm{\Phi}_y(\bm{x}_{*})
    \end{pmatrix} \begin{pmatrix}
        \bm{w}_u \\
        \bm{w}_v
    \end{pmatrix}\right\vert\right\vert_2^2 = \vert\vert \bm{D}_\nabla(\bm{x}_{*}) \bm{w}_U \vert\vert\vert_2^2 \,.
    \label{eq:rbf_penalty}
\end{equation} The penalty is only applied in the training points $\bm{x}_{*}$ but it could also be set in any point.

As linear constraints $\mathcal{H}$ in \eqref{Augmented}, we consider three terms. The first one further enforces the divergence-free condition in some specific points $\bm{x}_{\nabla}$; the associated linear constraint is similar to \eqref{eq:rbf_penalty} and reads:
\begin{equation}
    \mathcal{H}_\nabla = \begin{pmatrix}
        \bm{\Phi}_x(\bm{x}_{\nabla}) & \bm{\Phi}_y(\bm{x}_{\nabla})
    \end{pmatrix} \begin{pmatrix}
        \bm{w}_u \\
        \bm{w}_v
    \end{pmatrix} = \bm{D}_\nabla(\bm{x}_{\nabla}) \bm{w}_U\,.
\end{equation}

The constraint points $\bm{x}_\nabla$ need not coincide with the data points.
The second constraint is on the curl, which we impose on a set of points $\bm{x}_{\omega}$ and which reads as follows:

\begin{equation}
    \mathcal{H}_\omega = \begin{pmatrix}
        -\bm{\Phi}_y(\bm{x}_{c}) & \bm{\Phi}_x(\bm{x}_{c})
    \end{pmatrix} \begin{pmatrix}
        \bm{w}_u \\
        \bm{w}_v
    \end{pmatrix} = \bm{D}_\omega(\bm{x}_{\omega}) \bm{w}_U\,.
\end{equation}

Finally, the last constraint is on the boundary conditions. Both Dirichlet and Neuman conditions are possible (see \cite{Sperotto2022a}), but for this exercise, we only use the former. The linear constraint for the Dirichlet conditions sets:

\begin{equation}
    \mathcal{H}_D = \begin{pmatrix}
        \bm{\Phi}(\bm{x_{D}}) & \bm{0} \\
        \bm{0} & \bm{\Phi}(\bm{x_{D}})
    \end{pmatrix} \begin{pmatrix}
        \bm{w}_u \\
        \bm{w}_v
    \end{pmatrix} - \begin{pmatrix}
        \bm{c}_{Du} \\
        \bm{c}_{Dv}
    \end{pmatrix} = \bm{D}(\bm{x}_D) \bm{w}_U - \bm{c}_D\,,
\end{equation}
    
Assembling all terms in the augmented cost function \eqref{Augmented} leads to:

\begin{equation}
\label{Augmented_2}
\begin{split}
    \mathcal{A}(\bm{w}_U,\bm{\lambda}) =& \mathcal{J}(\bm{w}_U) + \alpha \vert\vert \bm{D}_\nabla(\bm{x}_{*}) \bm{w}_U \vert\vert\vert_2^2+ \bm{\lambda}^T_\nabla [\bm{D}_\nabla(\bm{x}_{\nabla}) \bm{w}_U] + \\
    &\bm{\lambda}^T_C [\bm{D}_C(\bm{x}_{C}) \bm{w}_U] + \bm{\lambda}^T_D [\bm{D}(\bm{x}_D) \bm{w}_U - \bm{c}_D]
\end{split}
\end{equation}

The careful choice of quadratic cost function, quadratic penalty, and linear constraints, together with the linearity of the RBF regression, allows for an analytic solution of the minimization of \eqref{Augmented_2}. Setting to zero the gradient with respect to both the weights and the Lagrange multipliers leads to the following linear system:

\begin{equation}
    \begin{pmatrix}
        \bm{A}_U & \bm{B}_U \\
        \bm{B}^T_U & \bm{0}
    \end{pmatrix} \begin{pmatrix}
        \bm{w}_U \\
        \bm{\lambda}_U
    \end{pmatrix} = \begin{pmatrix}
        \bm{b}_{U,1} \\
        \bm{b}_{U,2}
    \end{pmatrix}\,,
\end{equation}where the individual terms are defined as:

\begin{subequations}
    \begin{align}
        \bm{A}_U &= 2 \bm{\Phi}_U^T(\bm{x}_{*}) \bm{\Phi}_U(\bm{x}_{*}) + 2 \alpha \bm{D}_\nabla^T(\bm{x}_{*}) \bm{D}_\nabla(\bm{x}_{*}) \\
        \bm{B}_U &= (\bm{D}_\nabla^T(\bm{x}_{\nabla}),\,\bm{D}_C^T(\bm{x}_{C}),\,\bm{D}^T(\bm{x}_{D})) \\
        \bm{b}_{U,1} &= 2 \bm{\Phi}_U^T \bm{U} \\
        \bm{b}_{U,2} &= (\bm{0},\, \bm{0},\, \bm{c}_D)^T \\
        \bm{\lambda}_U &= (\bm{\lambda}_\nabla,\, \bm{\lambda}_C,\, \bm{\lambda}_D)\,.
    \end{align}
\end{subequations}

This linear system is solved with Cholesky factorizations and Schur complements, as described in \cite{Sperotto2022a}. In the provided SPICY toolbox, this is carried out in one command line.  We use the noisy training data as input to our model and then use the super-resolution property to recompute the solution on a regular grid.

\begin{figure}[htpb]
\center
 \includegraphics[width=0.99\textwidth]{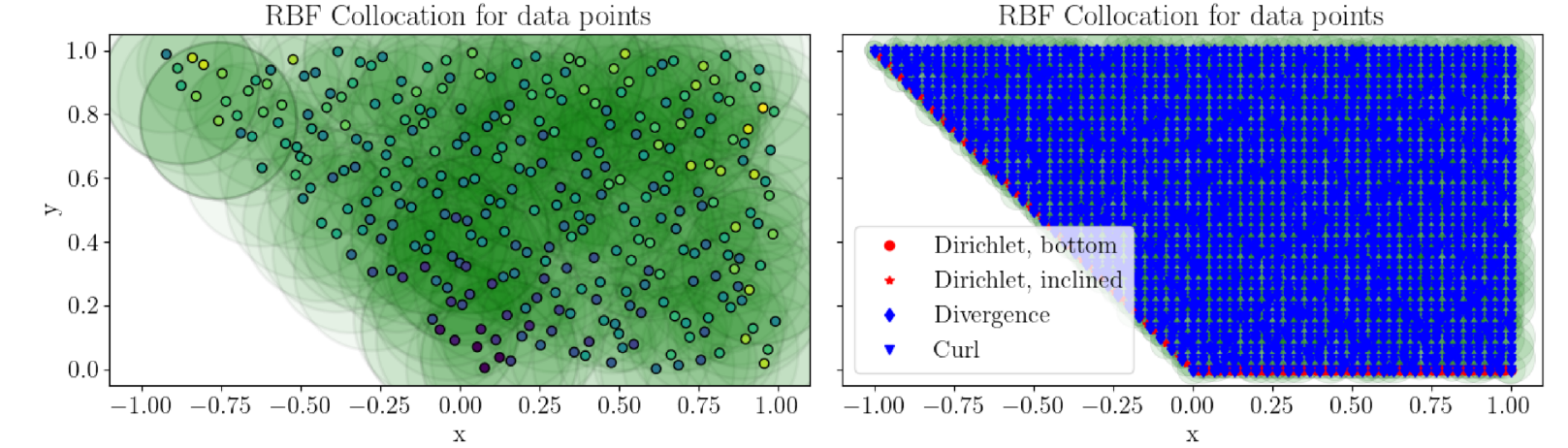}
    \centering
    \caption{Training data and RBFs collocation from K-means clustering (left) and constraints and RBFs in constraint points (right).}
    \label{fig:ex1_collocation_points}
\end{figure}

We include the previously mentioned physical priors at different points. The hard constraints of zero curl and zero divergence are \textit{both} applied in the training points $\bm{x}_{*}$, and the prediction points $\bm{x}_{**}$. This serves illustrative purposes for this idealized potential flow test case since the zero curl conditions in real experiments could only be applied far away from walls. Nevertheless, it helps illustrate how physical constraints enable extremely accurate regression and super-resolution fields. 
We take Gaussian RBFs from eq 14 from Chapter 2 as basis functions. The collocation points of the basis are found through K-means clustering, see \cite{Sperotto2022a} for details. Furthermore, we place an RBF in every point where we have a constraint because the constraints remove degrees of freedom from the regression. The resulting collocation points with the raw velocity data are shown on the left-hand side of Figure \ref{fig:ex1_collocation_points}, while the right-hand side shows the collocation points due to constraints and the constraints themselves.

\begin{figure}[htpb]
\center
 \includegraphics[width=0.99\textwidth]{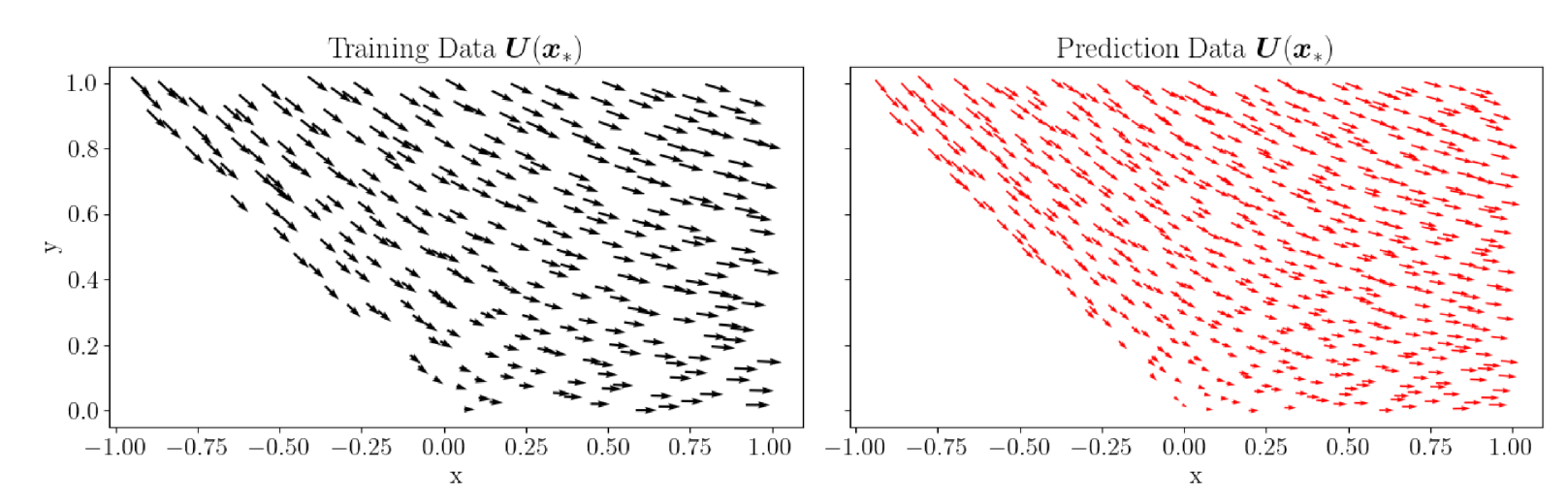}
    \centering
    \caption{Resulting quiver plot after physics-constrained regression. The velocity field in the training data (left) and velocity field are on a regular grid due to super-resolution (right).}
    \label{fig:ex1_result_quiver_plot}
\end{figure}

Figure~\ref{fig:ex1_result_quiver_plot} shows the reconstructed velocity field for both the training locations (left) and a dense prediction grid (right). Compared to Figure~\ref{fig:ex1_training_data}, the field is now smooth and outliers have been removed. The super-resolution capability allows us to visualize the solution on an arbitrarily fine grid, as illustrated on the right.

\begin{figure}[htpb]
\center
 \includegraphics[width=0.99\textwidth]{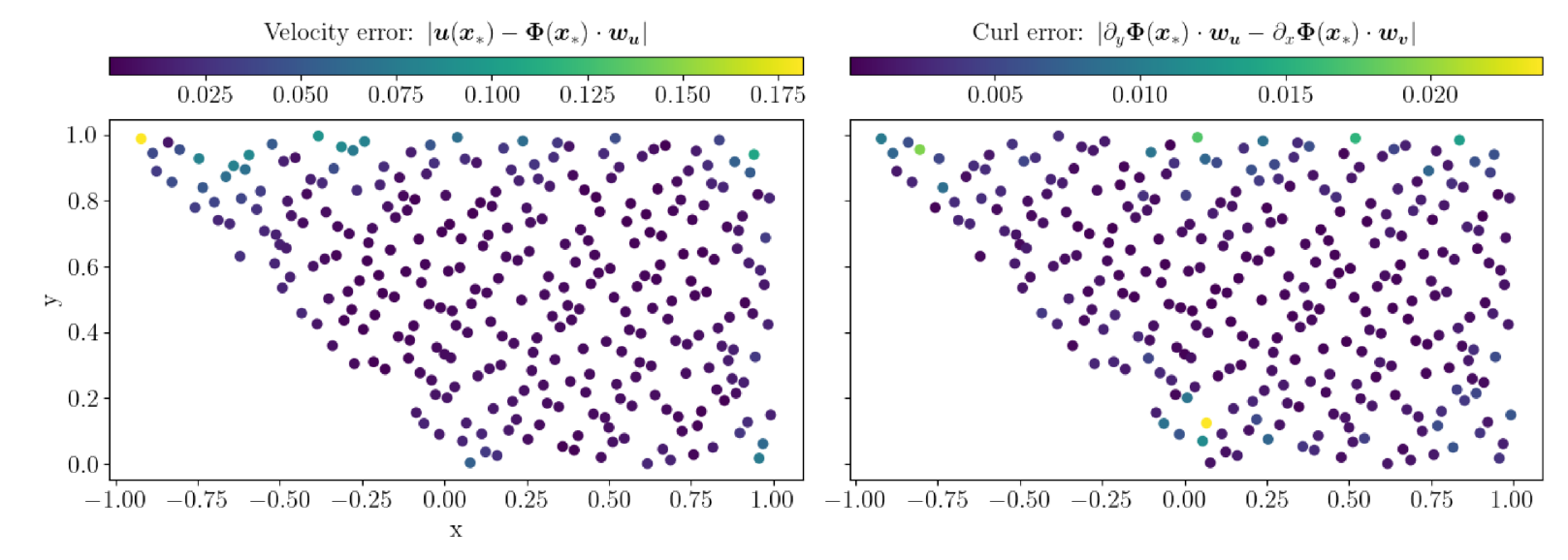}
    \centering
    \caption{Absolute error between RBF prediction and ground truth (left) and absolute error in the analytical curl field (right). }
    \label{fig:ex1_result_u_curl_training}
\end{figure}

A closer inspection of the pointwise errors is provided in Figure~\ref{fig:ex1_result_u_curl_training}, which shows the difference between the reconstructed velocity field and the ground truth at the training locations. The largest errors occur near the boundaries, where fewer data points are available; this is particularly noticeable along the top edge. The same trend is visible in the curl field shown on the right of the figure. Note that this curl field is obtained analytically by differentiating the basis matrix $\bm{\Phi}$. Although zero-curl constraints were enforced at the training points, the curl in the reconstructed field is not identically zero, and the divergence is likewise non-zero. This is a consequence of the regularization applied to the linear system to ensure numerical stability. Reducing the strength of the regularization would decrease the curl error but would do so at the expense of increasing the velocity reconstruction error.

\begin{figure}[htpb]
\center
 \includegraphics[width=0.94\textwidth]{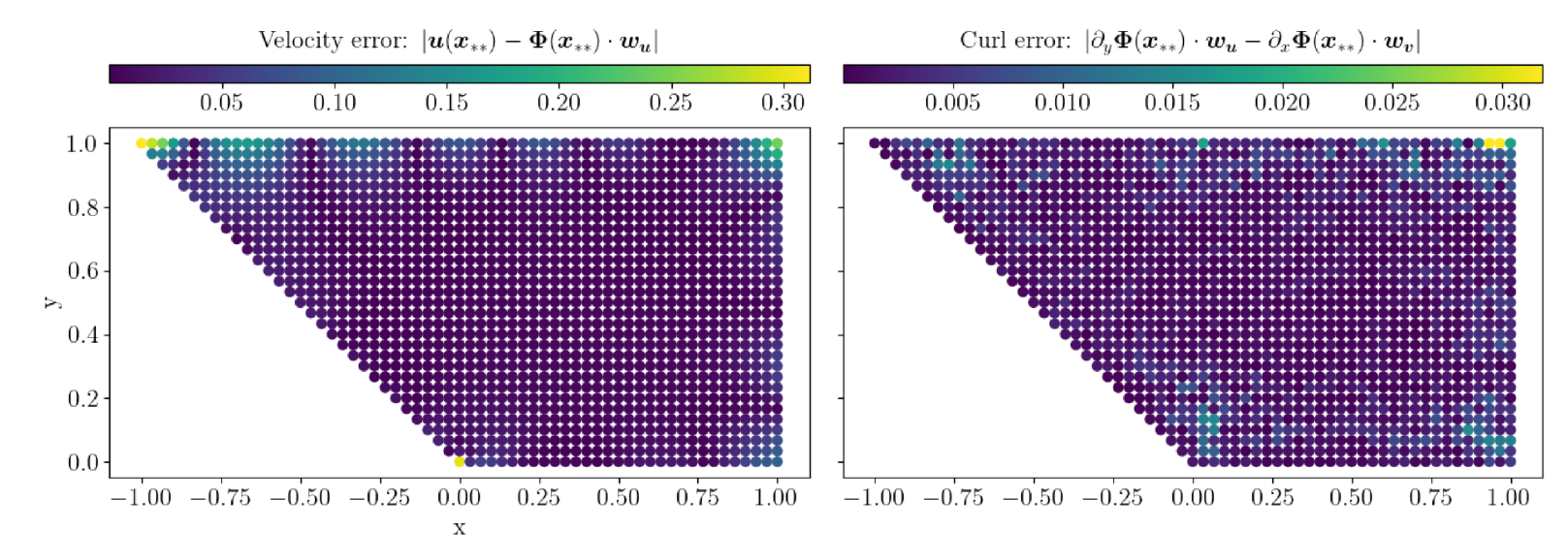}
    \centering
    \caption{Resulting errors in the prediction points. Absolute error between RBF prediction and ground truth (left) and absolute error in the analytical curl field. The error landscape is similar to the training data, and the model does not overfit.}
    \label{fig:ex1_result_u_curl_prediction}
\end{figure}

The same observations hold for the solution on the prediction points, the regular grid, which is shown in Figure \ref{fig:ex1_result_u_curl_prediction}. The errors in the velocity, curl, and divergence show the same distribution as before. This means that the regularization prevents overfitting, and the model yields accurate predictions, even on different prediction points.

One final remark is made regarding the relative velocity error. In both velocity components, the error is below 6\,\%, for both the training and the prediction data. This error is remarkable given the (unrealistic) 30\,\% noise in the ground truth and just 298 vectors for training. The results clearly show the advantages of physical constraints, which lead to small regression errors, even for very noisy and sparsely seeded datasets.

\section[Problem 2: Turbulence Modeling. What is the Reynolds Stress?]{Problem 2: Turbulence Modeling} \label{ch_3_sec_2}

\subsection{General Context}\label{ch_3_sec_2_2_1}
Turbulence modeling is a fundamental and still unresolved challenge in fluid dynamics. Most flows of practical relevance are turbulent, exhibiting chaotic and seemingly unpredictable behavior. Direct simulation of such flows from the Navier--Stokes equations requires computational resources that are typically prohibitive. Consequently, turbulence is often described using statistical theories. However, modeling the statistics of turbulent flows is difficult because the governing equations are inherently \emph{unclosed}. The two most common approaches are Large Eddy Simulation (LES) and Reynolds-Averaged Navier--Stokes (RANS), together with its unsteady counterpart (URANS). This brief introduction can only touch the surface of a vast literature, and the reader is referred to \cite{Tennekes,Benzi2023,Pope2000book} and to Chapter 4.

Both LES and RANS approaches solve for a \emph{filtered} representation of the velocity field. In LES, the filter is applied in wave-number space: the large, energy-containing scales are resolved, while the smaller scales are modeled. In URANS/RANS, the filtering is interpreted statistically, as an ensemble average over many realizations; the retained large scales are those that persist under this averaging. In either case, the flow variables are decomposed into a resolved part and a filtered part, and this decomposition is introduced into the Navier--Stokes equations. However, because of the nonlinear nature of the equations, the filtered-out fluctuations leave a residual term that influences the resolved dynamics. These terms represent the net effect of the unresolved scales: they remove energy from the resolved scales and appear in the momentum equations as additional stresses. In LES, they are known as \emph{subgrid-scale stresses}, while in URANS/RANS they are referred to as \emph{Reynolds stresses}. For an incompressible, isothermal flow, this results in ten unknown fields (the three velocity components, the pressure, and the six independent components of the additional stress tensor) but only four governing equations (one for mass conservation and three for momentum). This mismatch is the \emph{closure problem}: we must introduce a \emph{turbulence model} to relate the additional stresses to the resolved variables.

Focusing on the URANS/RANS formulation, the main classes of closure models are: (1) eddy-viscosity models, (2) Reynolds stress models, and (3) algebraic stress models. Eddy-viscosity models assume that the Reynolds stresses can be related to the mean velocity gradients through constitutive relations analogous to Newton’s law for viscous fluids. This introduces the concept of an \emph{eddy viscosity}, an effective viscosity representing the enhanced momentum diffusion caused by turbulence. Although this approach is mathematically convenient (reducing six unknown stress components to a single scalar field), it is often an oversimplification. Such models may include additional transport equations to compute the eddy viscosity. Reynolds stress models, in contrast, introduce one transport equation for each component of the Reynolds stress tensor, while algebraic models express the tensor as a linear combination of invariants of the mean velocity gradient.

These models typically contain empirically calibrated coefficients derived from canonical flows (e.g., free shear layers, boundary layers). Calibration is challenging due to the complexity of the governing equations, limited availability of high-quality experimental or numerical data, and strong sensitivity of predictions to model parameters. As a result, despite decades of research, no turbulence model has demonstrated robust generalization outside its calibration regime, and most models currently used in CFD codes are more than a decade old. Recent efforts and benchmarking activities in the community are summarized in the latest Turbulence Modeling Symposium\footnote{See \url{https://turbmodels.larc.nasa.gov/turb-prs2022.html}}, where coordinated comparison and testing campaigns were discussed.

Machine learning and advanced data assimilation techniques offer new perspectives on turbulence modeling (see \citet{Duraisamy2019}). However, these approaches are still in early development and face substantial challenges (see \citet{Spalart2023,Beck2021} and Chapter 4).

\begin{figure}[htpb]
\center
 \includegraphics[width=0.35\textwidth]{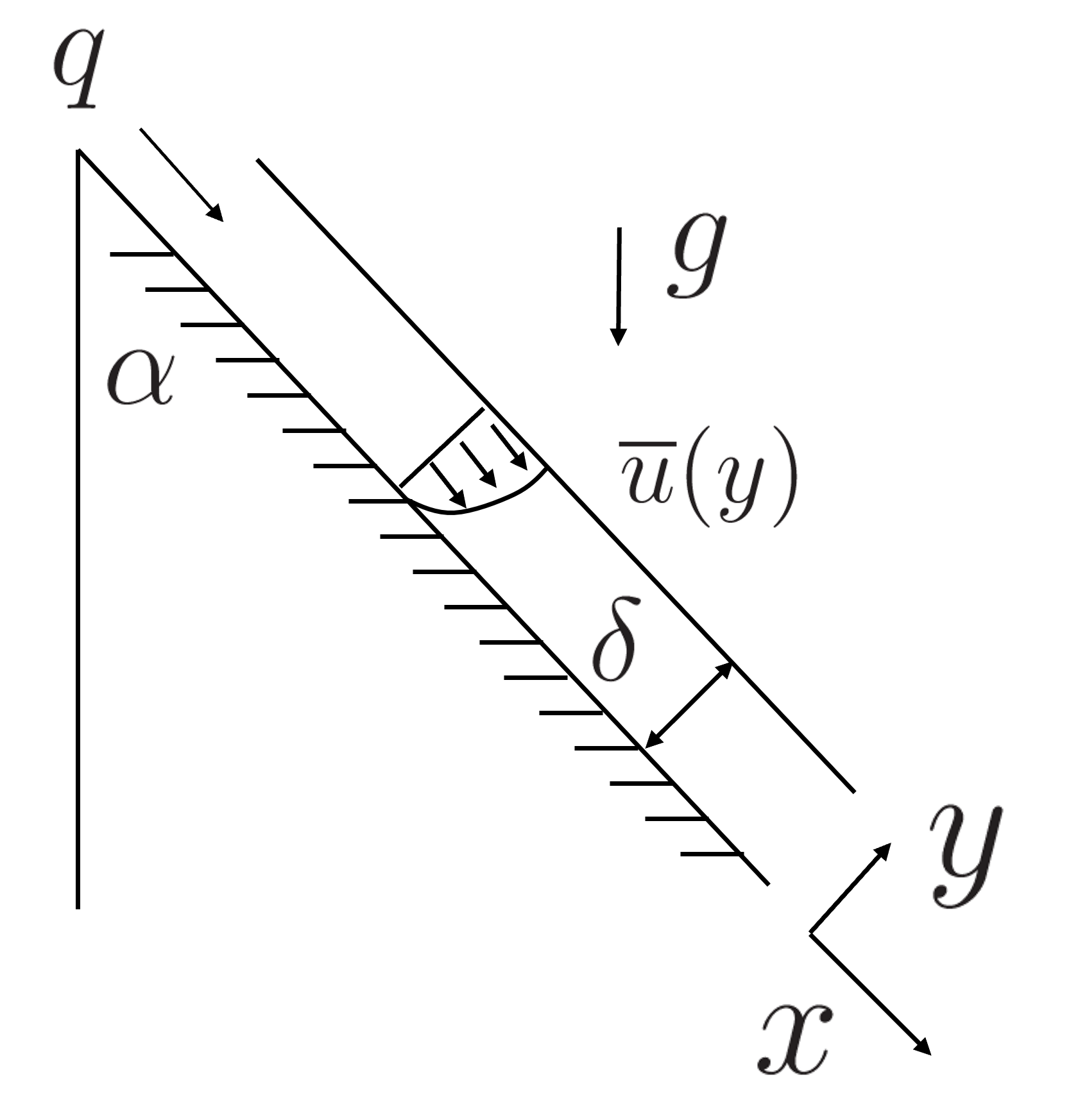}
    \centering
    \caption{Configuration investigated in this exercise: a turbulent liquid film falling along an inclined wall.}
    \label{liquidfilm}
\end{figure}

\subsection{Proposed Exercise}\label{ch_3_sec_2_2_2}

We consider the flow configuration in Figure \ref{liquidfilm}: a liquid film falls along an inclined wall. The flow rate per unit width $q$ and the average thickness of the film $\delta$ are known. The flow is turbulent, which produces fluctuations both in the velocity field and in the film thickness. The objective of the exercise is to identify the velocity profile $\overline{u}(y)$.

The flow is assumed stationary and turbulent, hence governed by the RANS momentum equation:
\begin{equation}
\overline{\bm{u}} \cdot \nabla \overline{\bm{u}} = -\bm{g} + \frac{1}{\rho} \nabla p + \frac{\mu}{\rho} \nabla^2 \overline{\bm{u}} - \nabla \cdot \overline{\bm{u}'\bm{u}'}
\label{mom_eq}
\end{equation} in which $\overline{\bm{u}'\bm{u}'}$ denotes the tensor of Reynolds stresses exerted by the unresolved turbulent structures on the mean flow. This quantity is unknown and requires to be modeled to integrate the velocity field and obtain its profile over the film thickness. Hence, the key point of the exercise is to provide a Reynolds stress closure for the problem under consideration, in other words, close eq. \eqref{mom_eq} with a turbulence model. 

In this exercise, we use a simple turbulence model to generate synthetic data and test whether a data-driven model can infer the underlying closure. More specifically, introducing the Boussinesq approximation to model the Reynolds stresses in terms of eddy viscosity $\mu_T$:
\begin{equation}
\overline{\bm{u}'\bm{u}'} = \frac{2}{3} k \bm{I} - \mu_T (\nabla \overline{\bm{u}} + (\nabla \overline{\bm{u}})^T )
\label{b_approx}
\end{equation} where $k$ is the turbulent kinetic energy and $\bm{I}$ denotes the identity matrix, the RANS equation \eqref{mom_eq} for the problem at hand simplifies to 

\begin{equation}
\frac{\partial }{\partial y} \left((\mu + \mu_T) \frac{\partial \overline{u}}{\partial y}\right) = -\rho g \cos{\alpha}\,.
\label{ss}
\end{equation}

The left-hand side is the shear stress $\tau$. Since $\tau=0$ at $y=\delta$, eq.\eqref{ss} can be integrated along $y$ to give:
\begin{equation}
(\mu + \mu_T) \frac{\partial \overline{u}}{\partial y} = -\rho g \cos{\alpha} (\delta-y)
\label{ss2}
\end{equation}

Eq.\eqref{ss2} can be conveniently re-scaled to highlight the dimensionless parameters governing this problem. Defining $\hat{y}=y/\delta$ and $\hat{u}=\overline{u}/U$, the dimensionless counterpart of \eqref{ss2} is:
\begin{equation}
\frac{d \hat{u}}{d \hat{y}} (1+ Re_T) = \frac{Re}{Fr^2} \cos{\alpha} (1-\hat{y})
\label{dimless_eq}
\end{equation} where $Re_T={\mu_T}/{\mu}$ and $Fr={U}/{\sqrt{g \delta}}$ govern the dynamics of the liquid film. In particular, $Re_T$ is the dimensionless eddy viscosity to be modeled to close eq. \eqref{dimless_eq} and compute the velocity profile. One of the easiest approximations of the turbulent shear stress is the Prandtl mixing length hypothesis, which reads as:

\begin{equation}
Re_T = \kappa \left |\frac{d \hat{u}}{d \hat{y}} \right| \hat{y}^2 Re
\label{prandtl_hypo}
\end{equation}
where $\kappa$ is the von Karman constant. Introducing \eqref{prandtl_hypo} in \eqref{dimless_eq} gives an analytical expression for the velocity gradient:
\begin{equation}
\frac{d \hat{u}}{d \hat{y}} = \frac{-1 \pm \sqrt{1 + 4 \kappa^2 Re^2/Fr^2 \cos{\alpha} \hat{y}(1-\hat{y})}}{2 \kappa^2 \hat{y}^2 Re}
\label{final_1d}
\end{equation}

\begin{figure}[t!]
%%----start of first subfigure----
\centering
\includegraphics[scale=0.5]{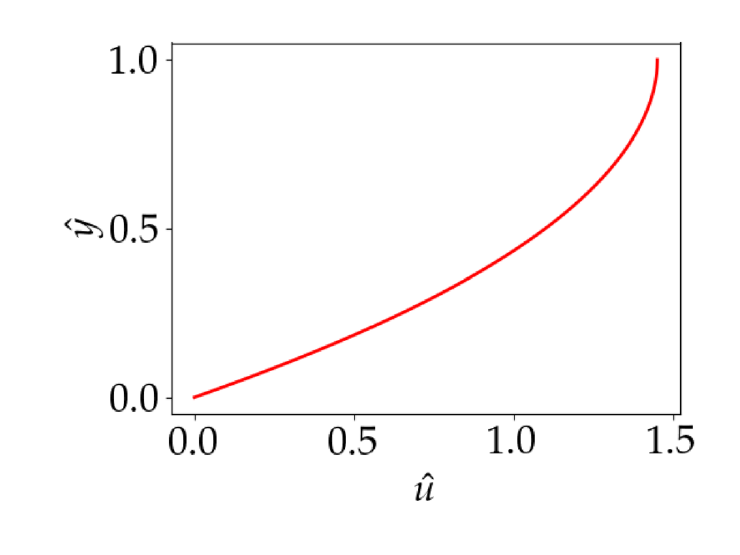}
\includegraphics[scale=0.5]{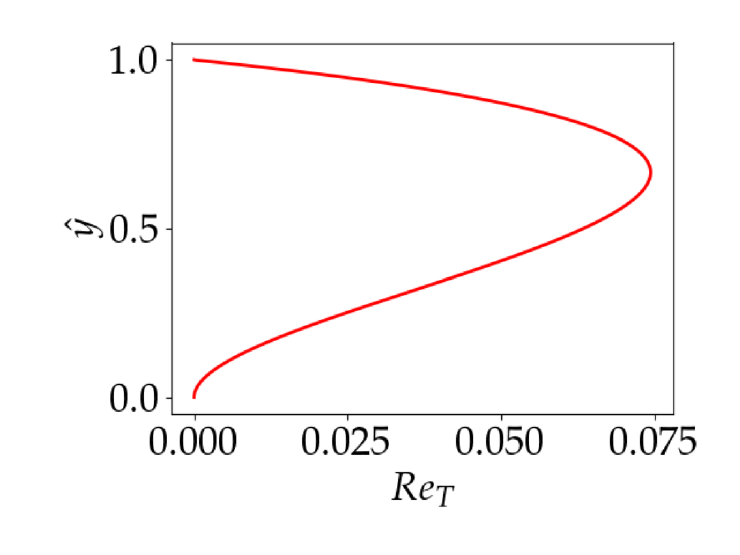}
\includegraphics[scale=0.5]{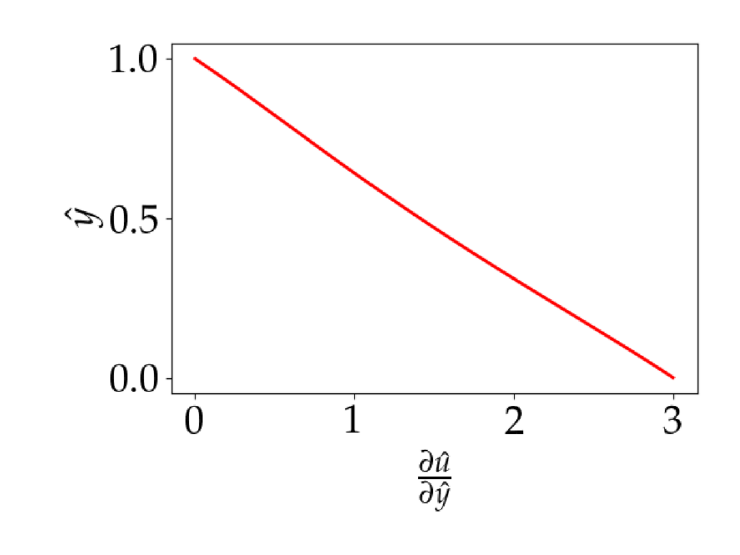}
\caption{Predictions of the analytical model used to generate the data.} %\cite{kawamura2000dns} (dashed lines) compared to those obtained by %applying eq. \ref{WE_all} (solid lines) for Pr=0.71.}
\label{profiles}%% label for entire figure
\end{figure}

Given the operating conditions $Re$, $Fr$, $\alpha$ and $\kappa$, \eqref{prandtl_hypo} can be integrated to completely characterize the flow. The Figures \ref{profiles} illustrate the velocity profile $\hat{u}(y)$, its derivative ${d \hat{u}}/{d \hat{y}}$ and the dimensionless eddy viscosity $Re_T$ obtained from Eq. \eqref{final_1d} with $Re=200$, $Fr=3$, $\kappa=0.4$ and $\cos{\alpha}=0.5$. 

\subsection{Methodology and Results}\label{ch_3_sec_2_2_3}
\begin{comment}
 We simulate a large set of data from \eqref{final_1d} and test the ability of a fully data driven model to retrieve the closure law in \eqref{prandtl_hypo}. This model is built using a fully connected feed-forward neural network.
\end{comment}
The purpose of this exercise is to illustrate how the sensitivity of a physical model affects the training of a data-driven model that aims to predict one of its closure terms. In practice, reference databases for turbulence modeling are typically obtained from experiments or high-fidelity simulations. Here, however, we construct a synthetic database using the law \eqref{final_1d} derived in the previous section. By sampling the parameters $Re$ and $\kappa$ over prescribed ranges, we generate a dataset of 1000 profiles of $\hat{u}$ and $Re_T$.

\begin{figure}[ht]
\center
 \includegraphics[width=1.0\textwidth]{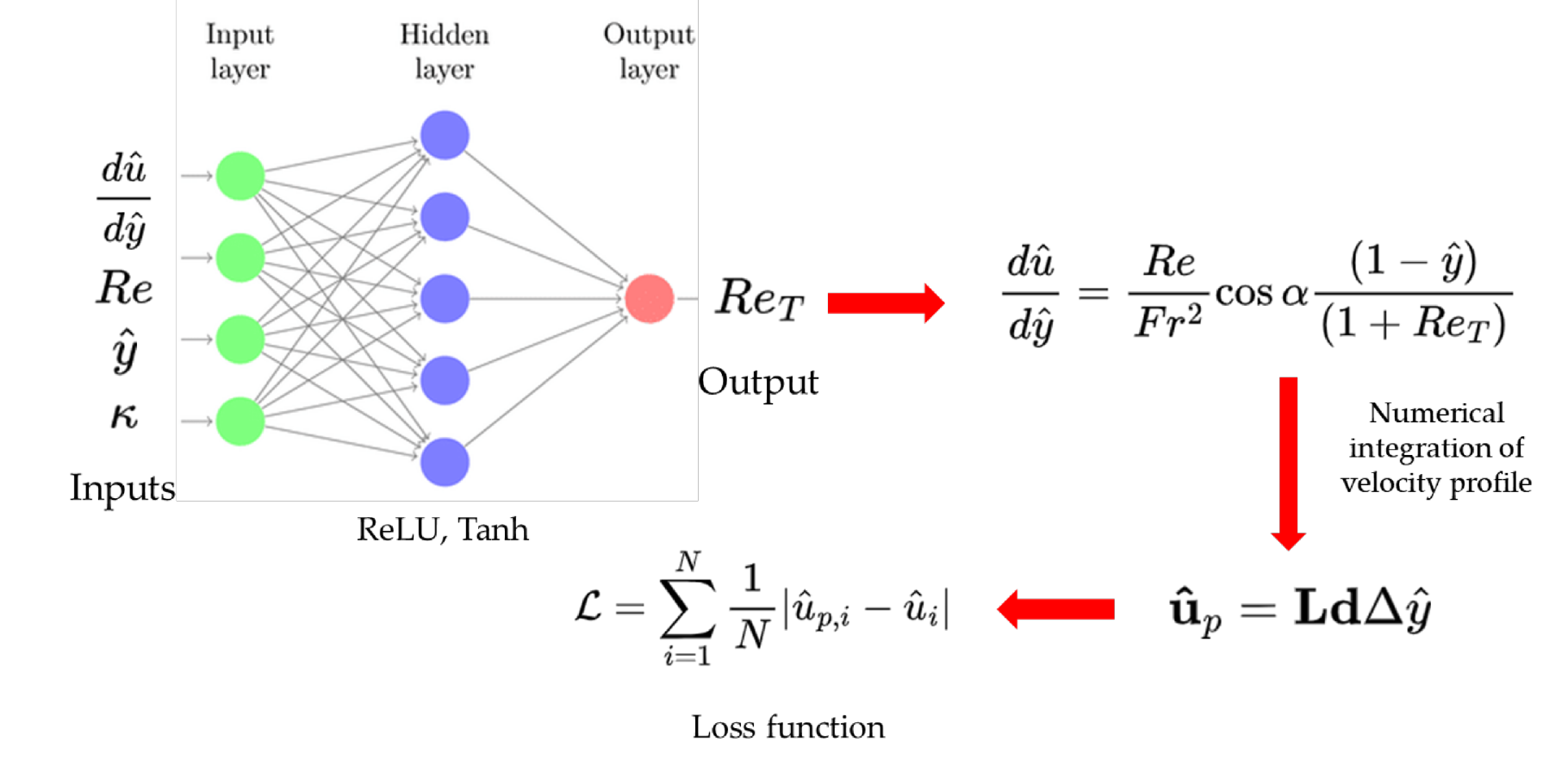}
    \centering
    \caption{Schematics of the hybrid architecture: the closure law is defined by an ANN which feeds a numerical solver for \eqref{final_1d}.}
    \label{Model_Architecture}
\end{figure}

The goal of the data-driven model is to predict the turbulent Reynolds number $Re_T(y)$ from a set of input quantities. For this exercise, we use the same inputs that appear in the mixing-length hypothesis \eqref{prandtl_hypo}, namely ${d\hat{u}}/{d\hat{y}}$, $\hat{y}$, $\kappa$, and $Re$. The ANN predictions are then supplied to a numerical solver, which integrates \eqref{final_1d} to recover the velocity profile.

The architecture of the regression model is shown in Figure~\ref{Model_Architecture}. The ANN consists of two hidden layers with Rectified Linear Unit (ReLU) and hyperbolic tangent activations. In principle, this setup could be made recursive: the ANN predicts $Re_T$, which determines a new ${d\hat{u}}/{d\hat{y}}$ that could be fed back into the network. However, for the purposes of this exercise, we restrict the model to a simple feed-forward configuration.

The network outputs the profile of $Re_T$, defining a mapping $N:\mathbb{R}^{4}\rightarrow\mathbb{R}$:
\begin{equation}
\label{wrong}
Re_T(\hat{y}) = N\!\left(\frac{d\hat{u}}{d\hat{y}}(\hat{y}),\, \kappa,\, Re,\, \hat{y}\right).
\end{equation}
This predicted quantity is substituted into \eqref{dimless_eq} to compute the velocity gradient ${d\hat{u}}/{d\hat{y}}$. Denoting the resulting discrete gradient values by the vector $\bm{d} := {d\hat{u}}/{d\hat{y}}(\hat{y})$, the velocity profile is then obtained through a straightforward numerical integration implemented as a matrix-vector product:
\begin{equation}
\bm{\hat{u}}_p = \bm{L}\,\bm{d}\,\Delta\hat{y},
\end{equation}
where $\bm{L}$ is a lower-triangular matrix of size $N\times N$, with $N$ the number of grid points.

The quality of the $\bm{\hat{u}}_p$ prediction is evaluated by a loss function that measures the mean absolute error between the reference and the computed velocity profiles:
\begin{equation}
\mathcal{L} = \sum_{i=1}^N \frac{1}{N} |\hat{u}_{p,i} -\hat{u}_{i} |
\label{loss}
\end{equation}

When the above regression model is trained, one would expect it to be able to reproduce the Prandtl mixing length hypothesis that generated the data. For a good picture of the performance of this regression model, it is important to estimate the epistemic uncertainty of the model, i.e. the effect of the uncertainty of the model parameters on the predictions. In the case of ANNs, the easiest strategy to estimate the uncertainty is to impose a dropout rate ($p=0.2$) at the network's hidden layers and keep it also in the prediction phase, as in \cite{gal2016dropout}. At each iteration, the effect of dropout is eliminating each node of the layer with a probability equal to the dropout rate. Note that this uncertainty quantification strategy is not quantitative (the choice of the dropout rate is arbitrary and would affect the confidence intervals), but qualitative, as it can highlight the regions of highest uncertainty in the input parameter space.  

\begin{figure}[ht]
\center
 \includegraphics[width=0.42\textwidth]{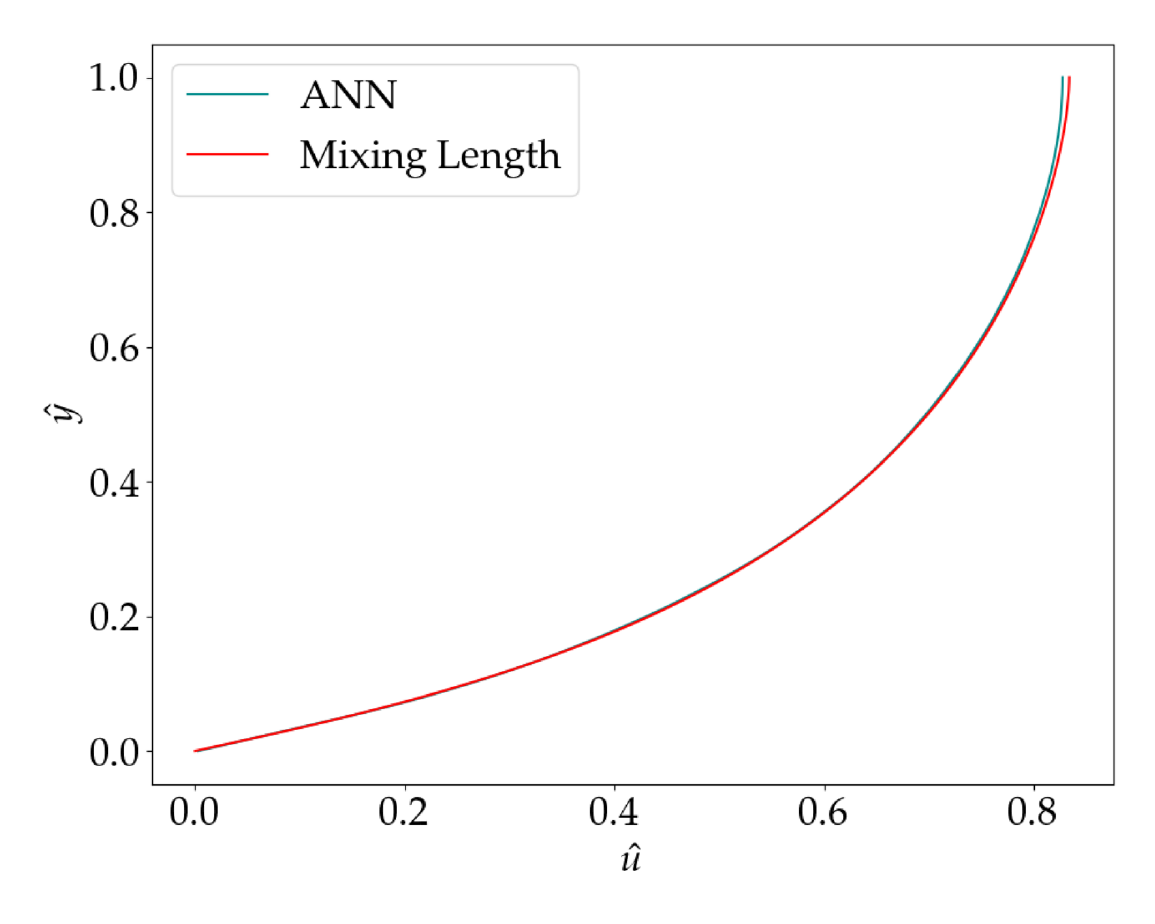} 
  \includegraphics[width=0.42\textwidth]{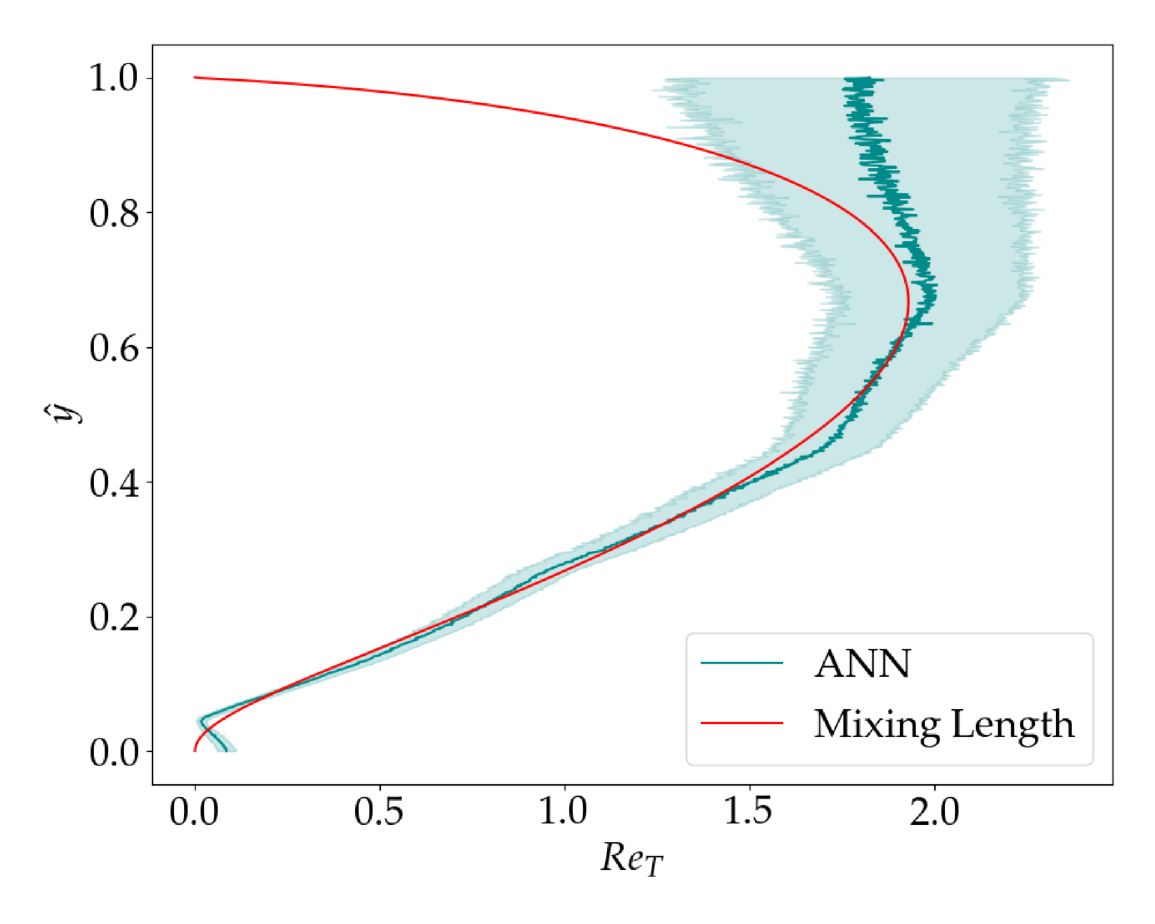}
    \centering
    \caption{Predictions of the velocity profile (top) and the turbulent Reynolds (bottom) for the problem sketched in Figure \ref{liquidfilm}.}
    \label{vel_profiles}
\end{figure}
Figure~\ref{vel_profiles} compares the ANN predictions for $\hat{u}(y)$ and $Re_T(y)$ with the analytical model derived in the previous section. The predicted velocity profile $\hat{u}(y)$ is highly accurate, with no visible uncertainty. The prediction of $Re_T(y)$, however, shows noticeable discrepancies, particularly near the interface.

Why does this happen? The issue is that the problem is \emph{ill-posed} in that region. The velocity profile (and therefore the loss function in Eq.~\eqref{loss}) is only weakly sensitive to the value of $Re_T(y)$ near the interface because the derivative ${d\hat{u}}/{d\hat{y}}$ is small there. In other words, the model receives almost no feedback from the loss in that region. The architecture in Figure~\ref{Model_Architecture} therefore encounters a vanishing-gradient problem: changes in the predicted closure have little effect on the velocity profile, which is the only quantity used to assess the model’s performance.

One way to address this is to incorporate physical knowledge directly into the model structure, following the first class of approaches discussed in Chapter 2. In this case, we know that the Reynolds stresses should vanish at both the wall and the interface\footnote{This assumption reflects the large viscosity ratio between liquid and gas when the gas is quiescent; it does not hold for high-speed gas shear over a liquid film.}. We can enforce this behavior by modifying the ANN output \eqref{wrong} to satisfy the correct boundary conditions by construction:
\begin{equation}
\label{corrected}
Re_T(\hat{y}) = \hat{y}\,(1-\hat{y})\,\mathcal{N}\!\left(\frac{d\hat{u}}{d \hat{y}}(\hat{y}),\,\kappa,\,Re,\,\hat{y}\right).
\end{equation}
This architectural change ensures the correct limiting behavior independently of the training data and improves the conditioning of the learning problem.

\begin{figure}[ht]
\center
  \includegraphics[width=0.42\textwidth]{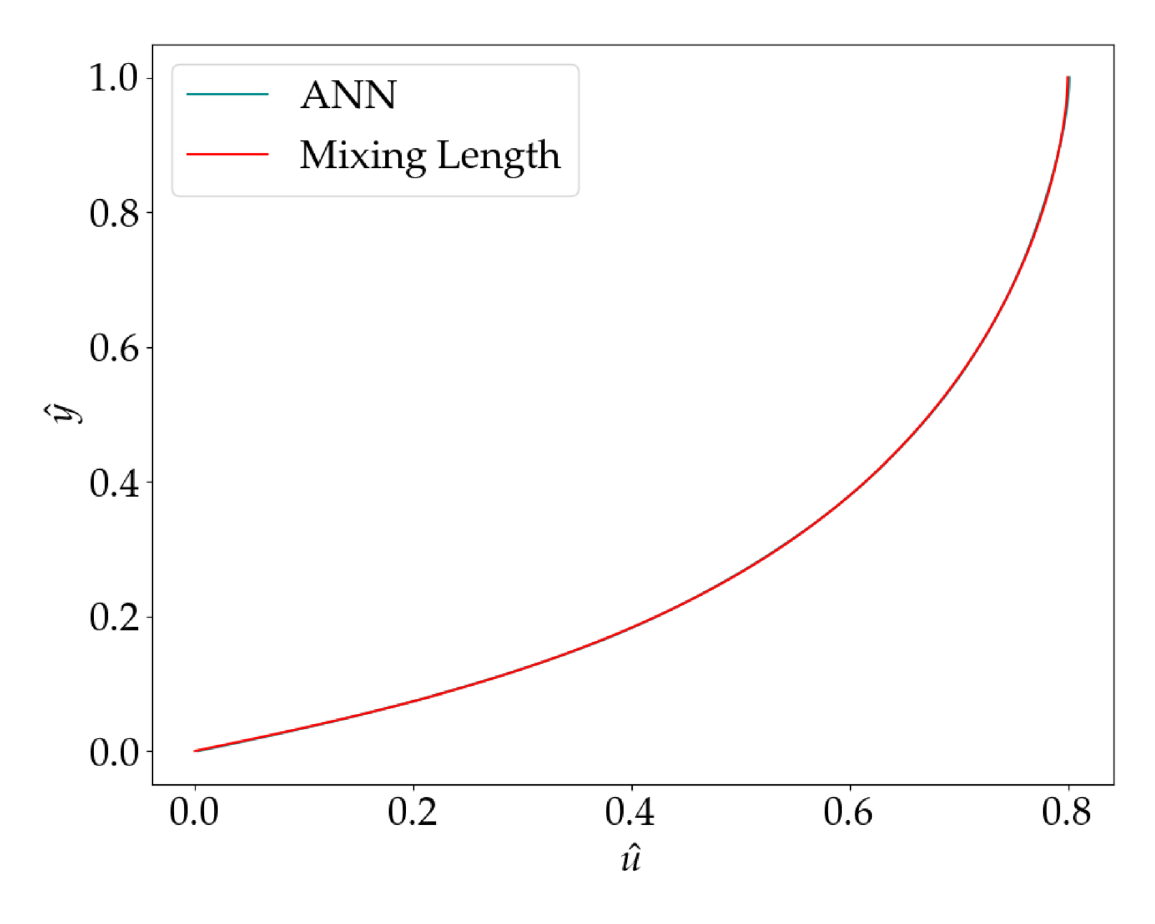}
 \includegraphics[width=0.42\textwidth]{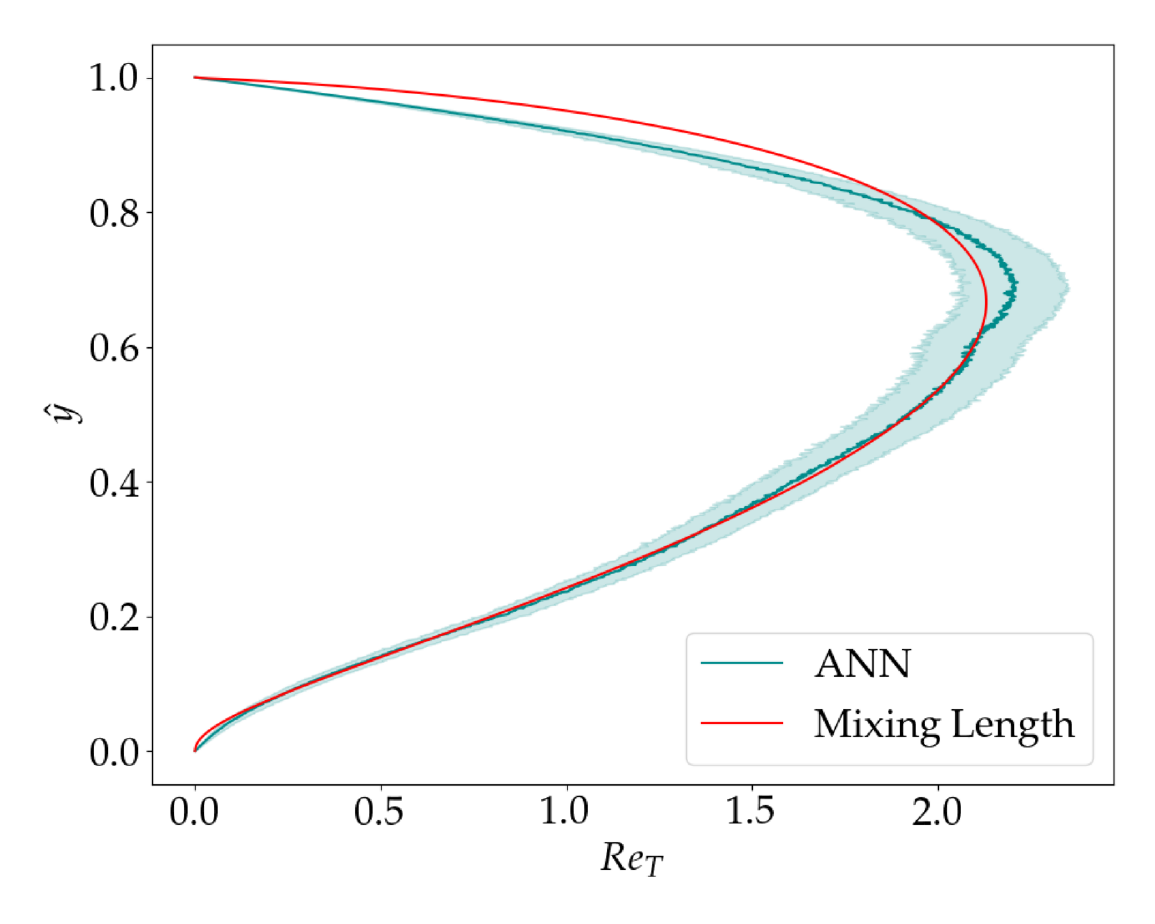}
    \centering
    \caption{Same as Figure \ref{vel_profiles}, but using the model in \eqref{corrected} instead of \eqref{wrong}.}
    \label{corrected_model}
\end{figure}

The training of the modified ANN gives the profiles of $\hat{u}(y)$ and $Re_T(y)$ depicted in Figure \ref{corrected_model}. The predictions of $Re_T$ are now consistent with the physics of the problem, and a small improvement of $\hat{u}(y)$ close to $\hat{y}=1$ is also visible. 

This simple exercise highlights the importance of architecturally constraining a data-driven model as much as possible to obey physical prior knowledge. Moreover, it gives an example of how the local ill-posedness of the regression, due to the vanishing of the network sensitivities, undermines the model performance. This is a very common issue in data-driven turbulence modeling because (1) errors in the turbulent statistics propagate to the mean velocity field through the resolution of transport equations, and (2) the turbulent statistics are usually determined by transport effects in vanishing gradient regions (e.g. zones of separation and reattachment) that cannot be represented with a single-point mapping.

\section{Problem 3: Forecasting and Control} \label{ch_3_sec_3}

\subsection{General Context}
\label{ch_3_sec_3_1}

We now move to a second exercise, conceptually similar to the previous one, in which a data-driven component is used to complement a physics-based model. However, we now consider a time-dependent setting—a dynamical system.

The task of improving or tuning a model based on observations of a system over time has a long history. With the rapid growth of sensor networks and IoT technologies, this problem has become increasingly relevant in the context of \emph{digital twins}: virtual replicas of physical systems that are continuously updated to reflect their evolving state and enable real-time prediction and control. Methods in this area vary depending on the available information, the level of uncertainty, and whether the goal is to maintain a physics-based representation (a ``white-box'' model) or replace it by a more flexible surrogate (a ``black-box'' model).

When the governing equations are assumed to be known but the system state is only partially observed, noisy, or high dimensional, the task is one of \emph{state estimation}. This is the realm of \textit{Data Assimilation} (DA) \citep{Asch2016,Bocquet2023}. Classical approaches include Variational Assimilation (4D-Var \cite{Talagrand1987,Ahmed2020}), which seeks an initial condition consistent with the observations, and Ensemble Kalman Filtering (EnKF \cite{Evensen2009,Routray2016}), which estimates a probabilistic representation of the system state.

In contrast, when the system is of relatively low dimension and the governing equations are known up to a set of unknown parameters, the task becomes one of \emph{parameter estimation} or \emph{nonlinear system identification}. Here, the objective is to infer model parameters (or, in some cases, the model form itself) from observed input–output trajectories. This can be interpreted as parametric regression in the context of dynamical systems. Classical examples include linear time-invariant and autoregressive models, as well as nonlinear representations such as Volterra series \citep{Schoukens2019}. System identification techniques have been widely developed in control theory, where external forcing dominates and uncertainties in initial conditions play a secondary role. Comprehensive reviews can be found in \cite{Nelles2001} and \cite{Suykens1996}.

Machine learning is increasingly influencing both data assimilation and system identification, with the dual aims of improving existing models and learning models directly from data. The need for real-time model adaptation is also growing within the Reinforcement Learning (RL) community \citep{sutton2018reinforcement}, where model-based approaches are gaining momentum \citep{chatzilygeroudis2019survey,luo2022survey,moerland2022modelbased} as a means to overcome the sample-efficiency limitations of classical model-free methods \citep{Werner2023,Pino2023}.

Recently, \citet{Schena2023} proposed a unified framework that blends reinforcement learning, system identification, and data assimilation to jointly train a digital twin of an engineering system and a control policy that optimizes its performance. In the present exercise, we explore a simplified version of this idea and introduce the main computational tool underlying variational data assimilation and optimal model-based control: the \emph{adjoint method} \citep{Stengel2012book}.

\subsection{The adjoint-based Identification Problem}\label{ch_3_sec_3_2}

We consider a parametric dynamical system of the form 

\begin{equation}
\label{sys}
 \begin{cases}
		\dot{\bm{s}}(t)&=f(\bm{s}(t),\bm{p})\\
		\bm{p}(t)&=\tilde{g}(\bm{s}(t);\bm{w}_p)\\		
		\bm{s}(0)&=\bm{s}_0\,.
	\end{cases}\,,
\end{equation} where $\bm{s}(t),\dot{\bm{s}}(t)\in\mathbb{R}^{n_s}$ are the state vector and its time derivative evolving according to the function 
$f: \mathbb{R}^{n_s\times n_p}\rightarrow \mathbb{R}^{n_s}$, $\bm{p}(t)\in\mathbb{R}^{n_p}$ is a set of unknown parameters, $\tilde{g}: \mathbb{R}^{n_s\times n_w}\rightarrow \mathbb{R}^{n_p}$ is a data-driven parametric closure function which depends on the parameters $\bm{w}_p\in\mathbb{R}^{n_w}$. 

We assume that the function $f$ is usually a physics-based model and the unknown parameter $\bm{p}$ are the necessary empirical coefficients (e.g., heat and mass transfer coefficients, aerodynamic coefficients, etc.) that must be predicted by the closure function $\tilde{g}$. Several applications of this framework are provided in \cite{Schena2023} and \cite{Marques2023}. The closure function could be a simple empirical correlation (e.g. \citet{Schena2023}) or a parametric model as complex as an ANN (e.g. in \cite{Marques2023}).

The scope of this exercise is to identify the set of parameters $\bm{w}_p$ while the system is running and data $\bm{s}^*(t)$ is being collected about its evolution. To measure the performances of the data assimilation, we define a differentiable and integral cost function:

\begin{equation}
\label{J_p}
\mathcal{J}_p(\bm{w})=\int^{T_0}_0 \mathcal{L}_p(\bm{s}(t;\bm{w}_p),\bm{s}^*(t_k)) dt\,,
\end{equation} with 

\begin{equation}
\mathcal{L}_p :=||\bm{s}(t;\bm{w})-\bm{s}^*(t_k)||^2_2\,.
\end{equation}
The adjoint method computes the gradient $d\mathcal{J}_p(\bm{w}_p)/d\bm{w}_p \in \mathbb{R}^{1\times n_p}$ without explicitly forming the sensitivities $d\bm{s}/d\bm{w}_p$. The idea is to exploit the fact that the state trajectories satisfy \eqref{sys} and introduce the augmented cost functional
\begin{equation}
\label{7_adj_J_1}
\mathcal{A}(\bm{w}_p)
=
\int_{0}^{T} \Big[\,\mathcal{L}_p(\bm{s}(t)) 
+ \bm{\lambda}_p(t)^{\mathsf{T}}\big(f(\bm{s}(t),\bm{w}_p)-\dot{\bm{s}}(t)\big)\Big]\, dt,
\end{equation}
where $\bm{\lambda}_p(t)\in\mathbb{R}^{n_s}$ is the vector of \emph{adjoint states}. The additional term inside the integral is identically zero for any $\bm{\lambda}_p(t)$ because it multiplies the system dynamics, which hold by construction. Therefore, $\bm{\lambda}_p(t)$ may be chosen freely, and we select it so that the gradient $d\mathcal{A}(\bm{w}_p)/d\bm{w}_p$ does not involve the sensitivities $d\bm{s}/d\bm{w}_p$. Furthermore, since the extra term vanishes on the solution of \eqref{sys}, we have $
{d\mathcal{A}(\bm{w}_p)}/{d\bm{w}_p}={d\mathcal{J}_p(\bm{w}_p)}/{d\bm{w}_p}.
$

Expanding the integral and using integration by parts for the term involving the time derivative of the states, \eqref{7_adj_J_1} can be written as:

 \begin{align}
\label{A2}
\mathcal{A}(\bm{w}_p) 
&= \int_0^T \left( \mathcal{L}_p + \bm{\lambda}_p^T f - \bm{\lambda}_p^T \dot{\bm{s}} \right)\, dt \notag\\
&= \int_0^T \left( \mathcal{L}_p + \bm{\lambda}_p^T f + \dot{\bm{\lambda}}_p^T \bm{s} \right)\, dt 
- \bm{\lambda}_p^T \bm{s} \Big|_0^T\,.
\end{align}  having omitted the functional dependencies for conciseness.

The gradient of \eqref{A2} reads:

\begin{align}
\label{dAdwp}
\frac{d\mathcal{A}}{d\bm{w}_p}
&= \int_0^T \biggl[
\frac{\partial\mathcal{L}_p}{\partial\bm{s}} \frac{d\bm{s}}{d\bm{p}}
+ \frac{\partial\mathcal{L}_p}{\partial\bm{p}}
+ \bm{\lambda}_p^T \left( \frac{\partial f}{\partial \bm{s}} \frac{d\bm{s}}{d\bm{p}} + \frac{\partial f}{\partial \bm{p}} \right)
+ \dot{\bm{\lambda}}_p^T \frac{d\bm{s}}{d\bm{p}}
\biggr] \frac{d\tilde{g}}{d\bm{w}_p}\,dt \notag\\
&\quad - \bm{\lambda}_p^T \frac{d\bm{s}}{d\bm{p}} \frac{d\tilde{g}}{d\bm{w}_p} \Big|_0^T\,.
\end{align}

Factoring all sensitivities $d\bm{s}/d\bm{p}$ one gets 
\begin{align}
\label{dAdwp_factored}
\frac{d\mathcal{A}}{d\bm{w}_p}
&= \int_0^{T_o}
\biggl[
\color{red}{
\left( \frac{\partial \mathcal{L}_p}{\partial \bm{s}} + \bm{\lambda}_p^T \frac{\partial f}{\partial \bm{s}} + \dot{\bm{\lambda}}_p^T \right)
}
\color{black}{
\frac{d\bm{s}}{d\bm{p}}
+
\frac{\partial \mathcal{L}_p}{\partial \bm{p}}
+ \bm{\lambda}_p^T \frac{\partial f}{\partial \bm{p}}
}
\biggr]
\frac{d\tilde{g}}{d\bm{w}_p} \, dt \notag\\
&\quad
\color{red}{
- \bm{\lambda}_p^T \frac{d\bm{s}}{d\bm{p}} \frac{d\tilde{g}}{d\bm{w}_p} \Big|_0^{T_o}
}
\color{black}{\, .}
\end{align}

Here is where one uses the liberty of picking $\bm{\lambda}_p$ arbitrarily so that the sensitivities are no longer relevant. One could pick the adjoint states such that

\begin{equation}
	\label{lambda}
	\begin{cases}
		\dot{\bm{\lambda}}_p(t)&=-\bigl(\frac{\partial \mathcal{L}_p}{\partial \bm{s}}\bigr)^T +  \bigl(\frac{\partial f}{\partial \bm{s}}\bigr) \bm{\lambda}_p\\
		\bm{\lambda}_p(T_o)&=\bm{0}\,.
	\end{cases}
\end{equation} 

So that then the gradient is 

\begin{equation}
	\frac{d\mathcal{J}_p}{d\bm{w}_p}(\bm{w}_p)= \int^{T_o}_0
	\biggl(\frac{\partial \mathcal{L}_p}{\partial \bm{p}}
+ \bm{\lambda}_p(t)^T \frac{\partial f}{\partial \bm{p}}\biggr)\frac{d\tilde{g}}{d\bm{w}_p} \,dt
\end{equation}

with:
\begin{equation}
\label{eq:adjoint_dims}
    \frac{\partial\mathcal{L}_p}{\partial \bm{p}}\in\mathbb{R}^{1\times n_p}\,, \quad
    \frac{\partial f }{\partial \bm{p}}\in\mathbb{R}^{n_s\times n_p}\,, \quad
    \frac{d \tilde{g}}{d \bm{w}_p}\in\mathbb{R}^{n_p\times n_w}\,,\quad 
    \bm{\lambda}_p\in\mathbb{R}^{n_s}\,.
\end{equation}

The adjoint states thus represent the sensitivity of the cost function to the dynamic of the system, computed proceeding backward in time from the terminal state $\boldsymbol{\lambda}(T_o)$. Importantly, this method does not scale with the number of parameters $n_p$ and allows for computing the gradient ${d\mathcal{J}_p}/{d\bm{w}_p}$ with a single backward pass. This makes it particularly appealing for optimization problems with a large number of parameters, as is the case, for instance, where $\tilde{g}(\bm{s},\bm{w}_p)$ is an ANN.

\subsection{Proposed exercise}
\label{ch_3_sec_3_3}

\begin{figure}[ht]
     \centering
     \begin{subfigure}[b]{0.45\linewidth}
     \centering
         \includegraphics[width=0.6\textwidth]{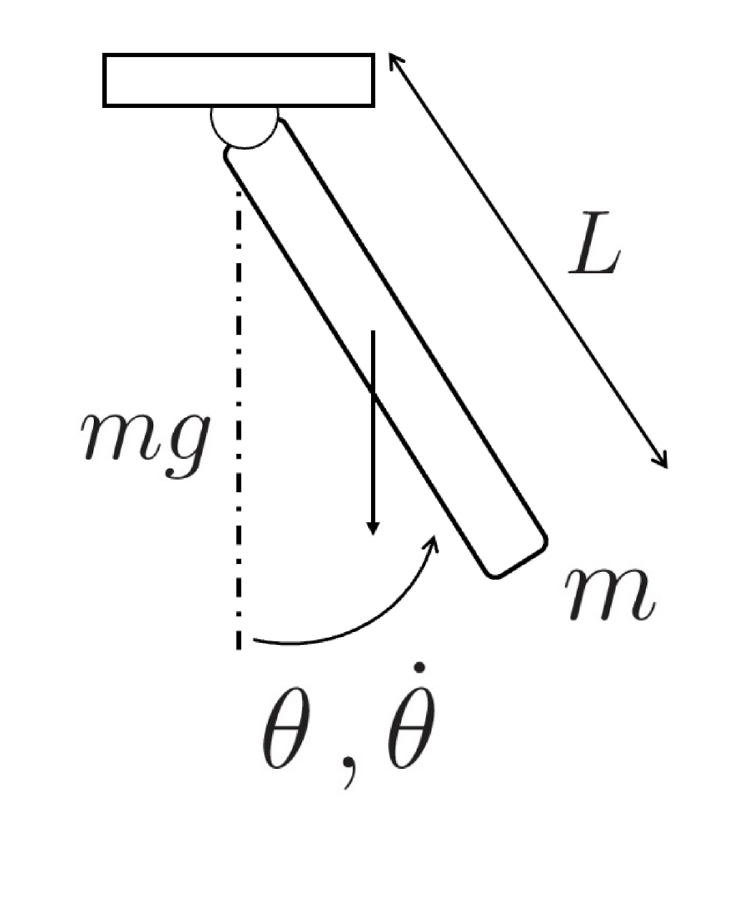}
         \caption{}
         \label{fig:pendulum_sketch}
     \end{subfigure}
     \hfill
     \begin{subfigure}[b]{0.47\linewidth}
     \centering
         \includegraphics[width=0.8\textwidth]{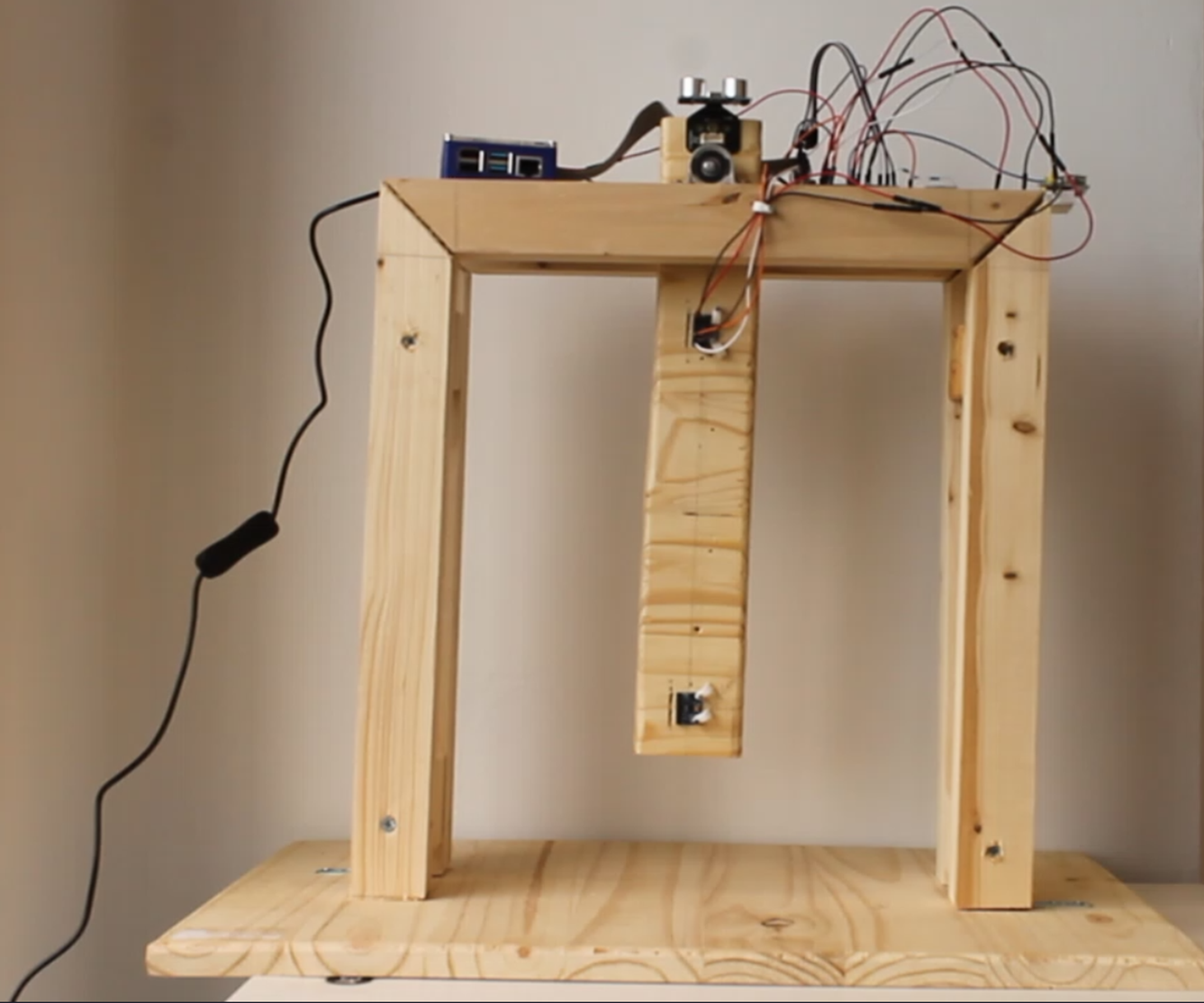}
         \caption{}
         \label{fig:pendulum_pict}
     \end{subfigure}
        \caption{Pendulum system investigated in this exercise. The goal is to identify the system parameters from sensor data while the system is running.}
        \label{fig:pendulum_sketches}
\end{figure}

In the proposed exercise, we employ the adjoint technique to identify the parameters of a simple nonlinear model, that is, a physical pendulum. The system is schematized in Figure \ref{fig:pendulum_sketch}. The data is measured from a simple experimental set-up, shown in Figure \ref{fig:pendulum_pict}. The setup is instrumented to sample the full state vector $\bm{s}(t)=[\theta(t),\dot{\theta}(t)]^T\,$, which evolves according to 
\begin{equation}
\label{eq:pendulum_sys_s}
 \begin{cases}
	\dot{s}_1 &=  \bm{s}_2  \\
	\dot{s}_2 &= -\mu \bm{s}_2 + \omega^2_n \sin(s_1)
 \end{cases} 
 \,.
\end{equation} 

The system state is sampled with a sampling frequency $f_s$=50 s, and one episode lasts $T=$10 s. The mass of the pendulum is $m=$120 g, and its length $L=$41 cm. The coefficient to be identified is the damping ($\approx\mu=$0.18 s$^{-1}$ and $\omega_n=$5.73 rad s$^{-1}$). For the simple problem at hand, there is no need for closure hence the mapping $\tilde{g}$ in \eqref{sys} is the identity function, that is $\bm{p}=\bm{w}=[\mu,\omega_n]$ and its Jacobian $d\tilde{g}/d\bm{w}$ is the identity matrix.

In this problem, we assume a fixed set of parameters, but these may generally be time-varying; the same approach would apply with minor modifications. The identification algorithm is illustrated in Algorithm \ref{algo:adjoint_pendulum}.

\begin{algorithm}
\caption{Adjoint Regression}\label{algo:adjoint_pendulum}
  \textbf{Input:} Real system trajectory ${\boldsymbol{s}(t)}$, set of possible initial conditions $\boldsymbol{s}_0$. Initial guess for $\bm{p} = \boldsymbol{w}_p^0$. \\
  \While{forever}{
    Collect the system observation within the episode: $\bm{s}^*(t_0)$, $\bm{s}^*(t_1)$, ... $\bm{s}^*(t_{n_t-1})$
    Given the estimate of the parameters $\bm{p}^{(i)}$, run a simulation on the digital twin to collect the predictions  $\bm{s}(t_0)$, $\bm{s}(t_1)$, ... $\bm{s}(t_{n_t-1})$ \\ 
    Evaluate performance and its gradients via the adjoint method: $\mathcal{J}_p^{(i)} , \partial_p \mathcal{J}^{(i)}$ \\ 
    Update parameters with the optimization method of choice:
    \begin{equation}
    \label{eq:w_p_evolution}
        \boldsymbol{w_p}^{(i+1)} = \boldsymbol{w_p}^{(i)} +\Delta \bm{w}_p\biggl(\frac{d\mathcal{J}}{d\bm{w}_p}\biggr) 
    \end{equation}
  }
\end{algorithm}

\begin{figure}
     \centering
     \begin{subfigure}[b]{0.45\textwidth}
         \centering
         \includegraphics[width=0.97\textwidth]{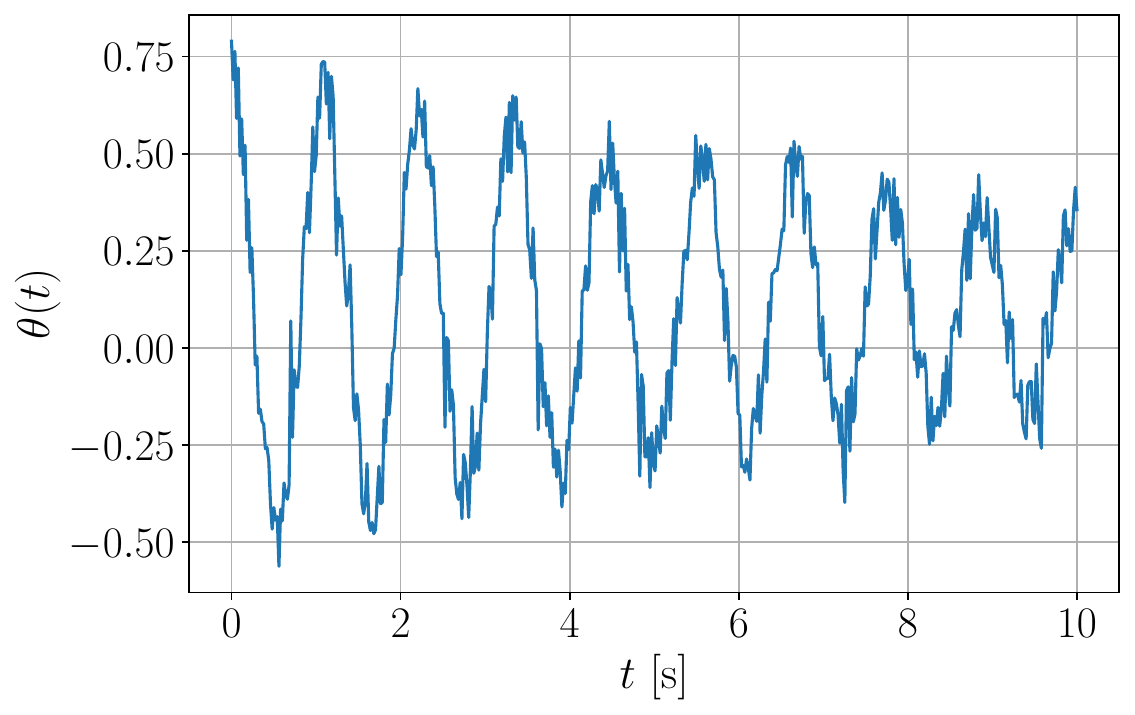}
         \caption{$\boldsymbol{s}_1(t) = \theta (t)$}
         \label{fig:y equals x}
     \end{subfigure}
     \begin{subfigure}[b]{0.43\textwidth}
         \centering
         \includegraphics[width=0.97\textwidth]{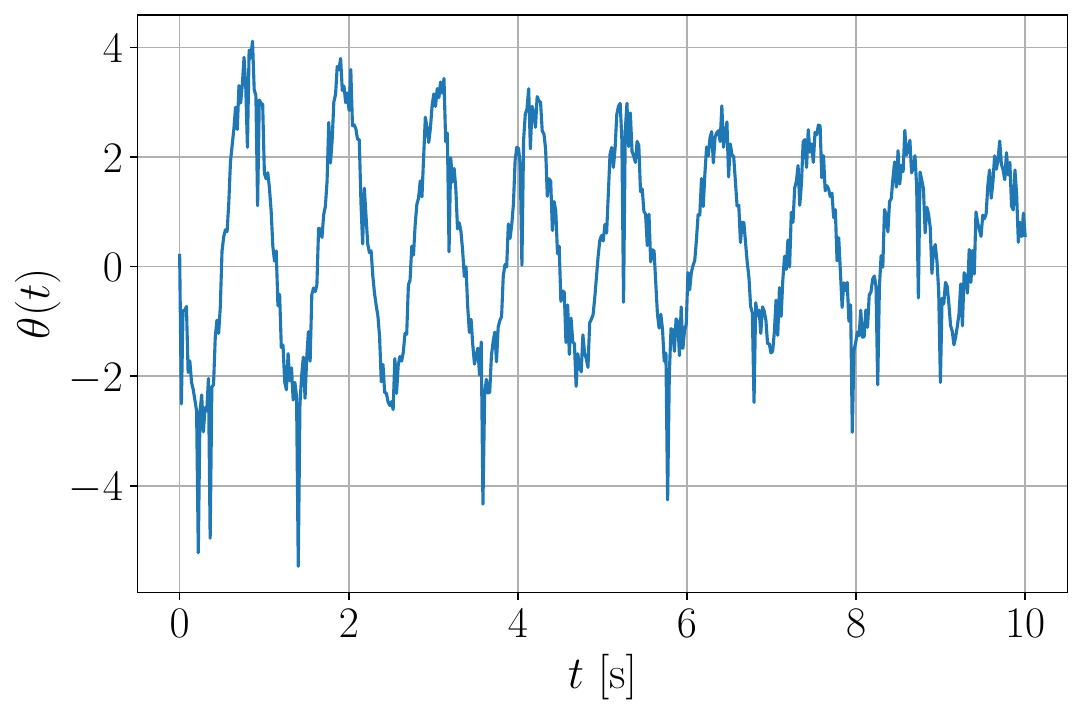}
         \caption{$\boldsymbol{s}_2(t) = \dot{\theta} (t)$}
         \label{fig:three sin x}
     \end{subfigure}
        \caption{Examples of sampled states during one of the episodes}
        \label{fig:pendulum_ex_trajectories}
\end{figure}

In this exercise, we simulate the sampling of $n=$8 episodes of the real system, and we seek to tailor a model that follows the real counterpart.  We use the ADAM optimizer introduced in Section \ref{ch_2_sec_2_1} with a learning rate of $\alpha=0.002$ and the default $\beta$ parameter. An example of sampled trajectory is shown in Figure \ref{fig:pendulum_ex_trajectories}, which shows some associated noise in the acquisition process. 

\begin{figure}[ht]
     \centering
     \begin{subfigure}[b]{0.45\linewidth}
         \centering
         \includegraphics[width=0.97\textwidth]{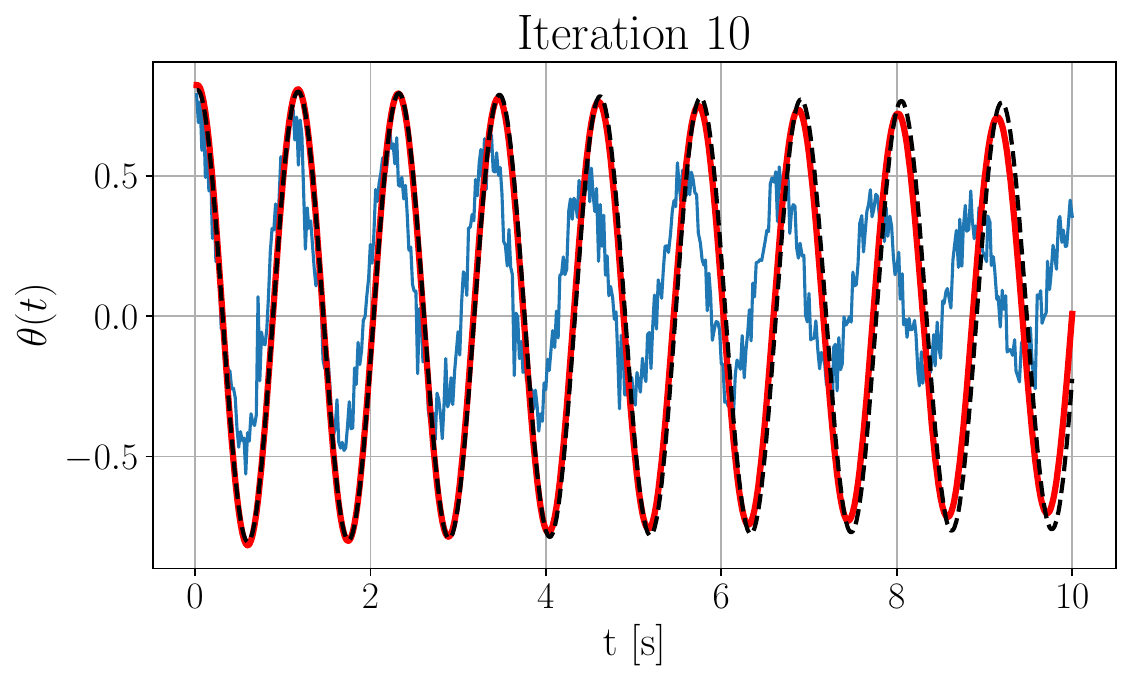}
         \caption{}
         \label{fig:ep_10_dyn}
     \end{subfigure}
     \hfill
     \begin{subfigure}[b]{0.42\linewidth}
         \centering
         \includegraphics[width=0.97\textwidth]{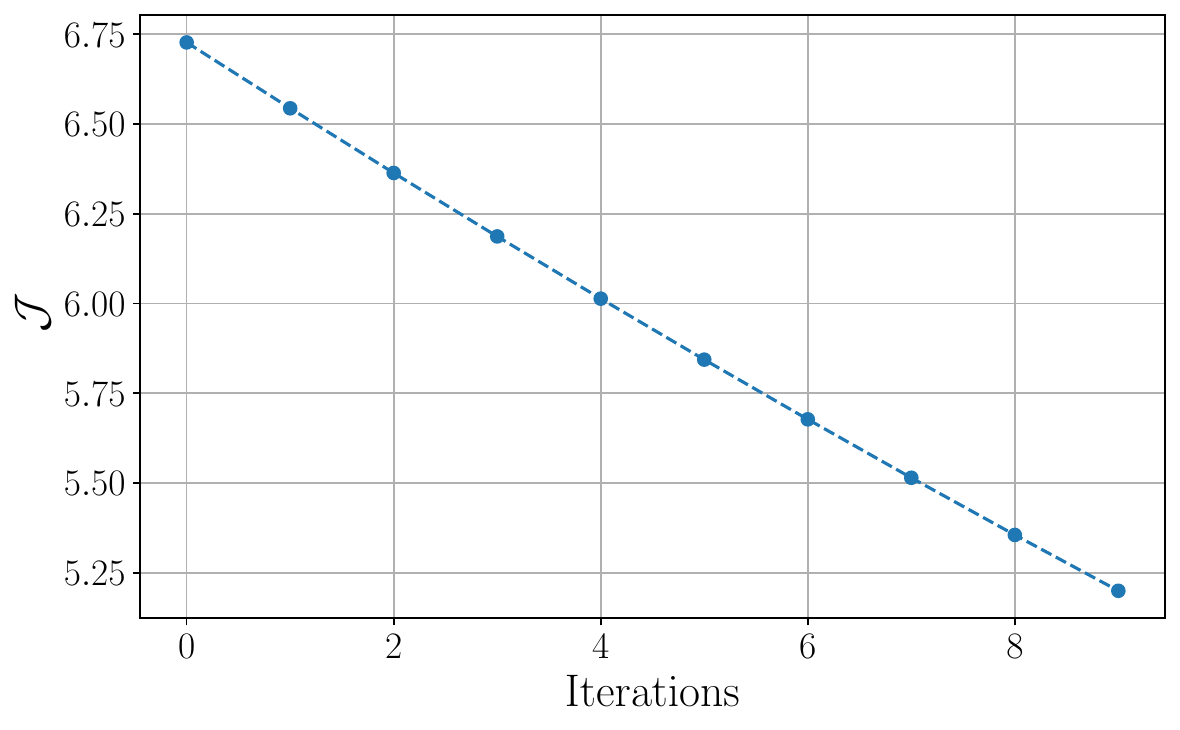}
         \caption{}
         \label{fig:ep_10_cost}
    \end{subfigure}

    \begin{subfigure}[b]{0.45\linewidth}
         \centering
         \includegraphics[width=0.97\textwidth]{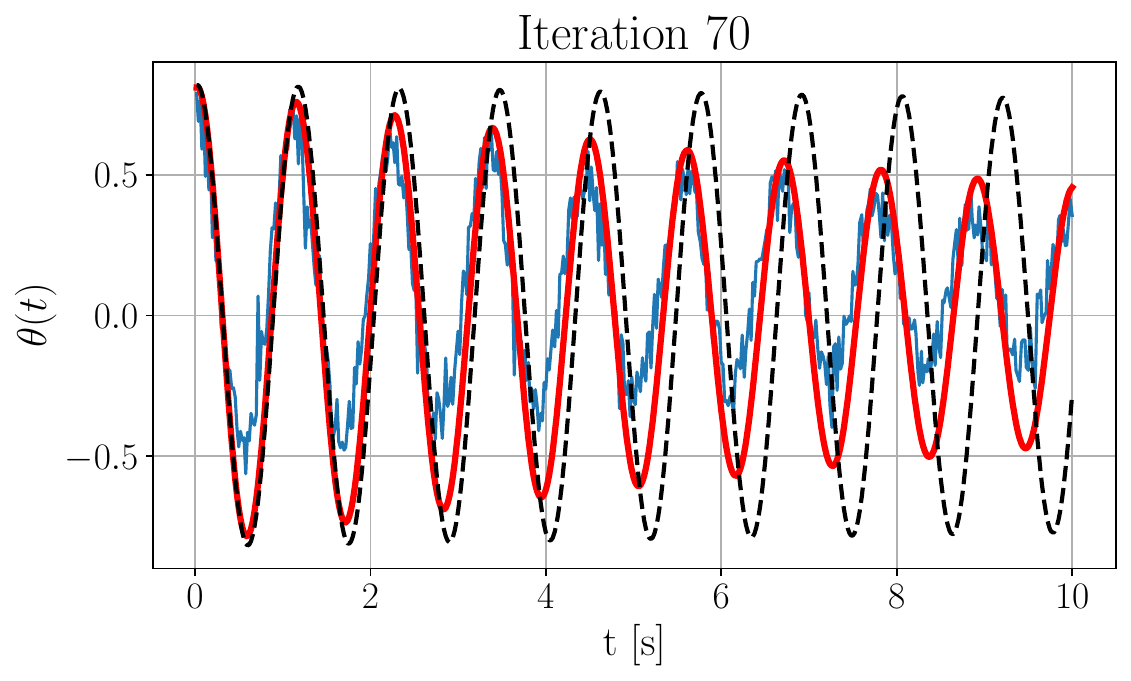}
         \caption{}
         \label{fig:ep_100_dyn}
     \end{subfigure}
     \hfill
     \begin{subfigure}[b]{0.42\linewidth}
         \centering
         \includegraphics[width=0.97\textwidth]{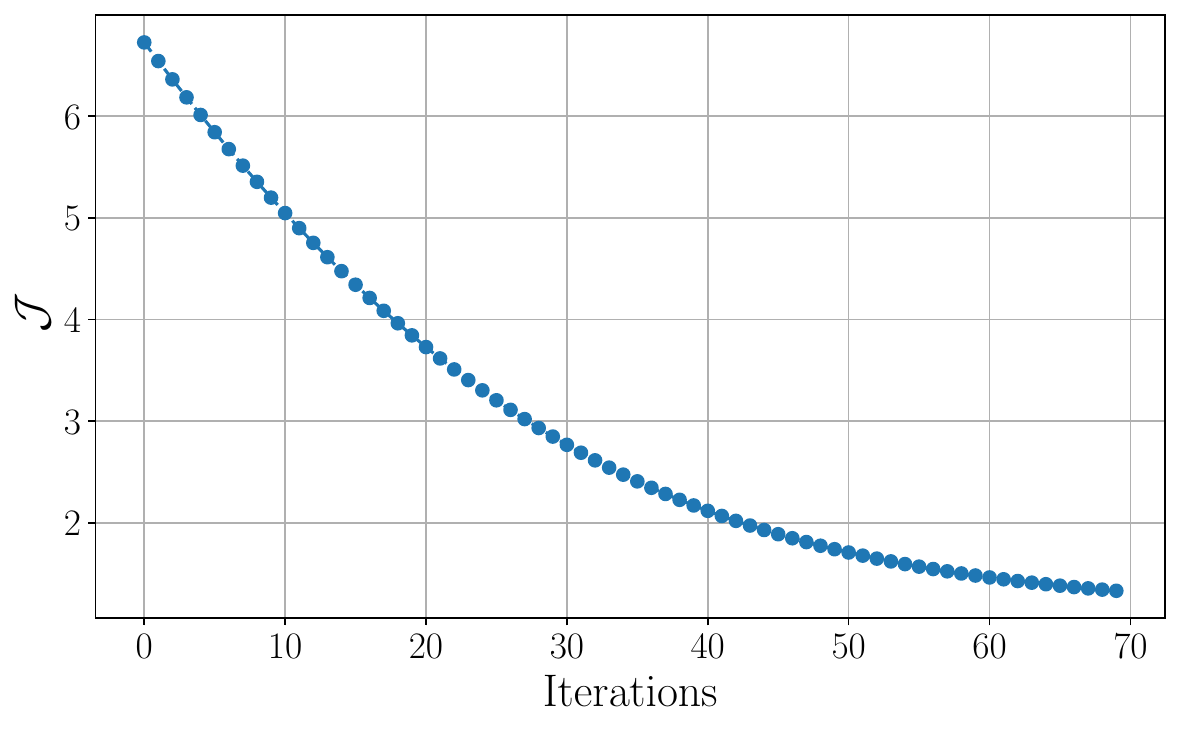}
         \caption{}
         \label{fig:ep_100_cost}
    \end{subfigure}
    
    \begin{subfigure}[b]{0.45\linewidth}
         \centering
         \includegraphics[width=0.97\textwidth]{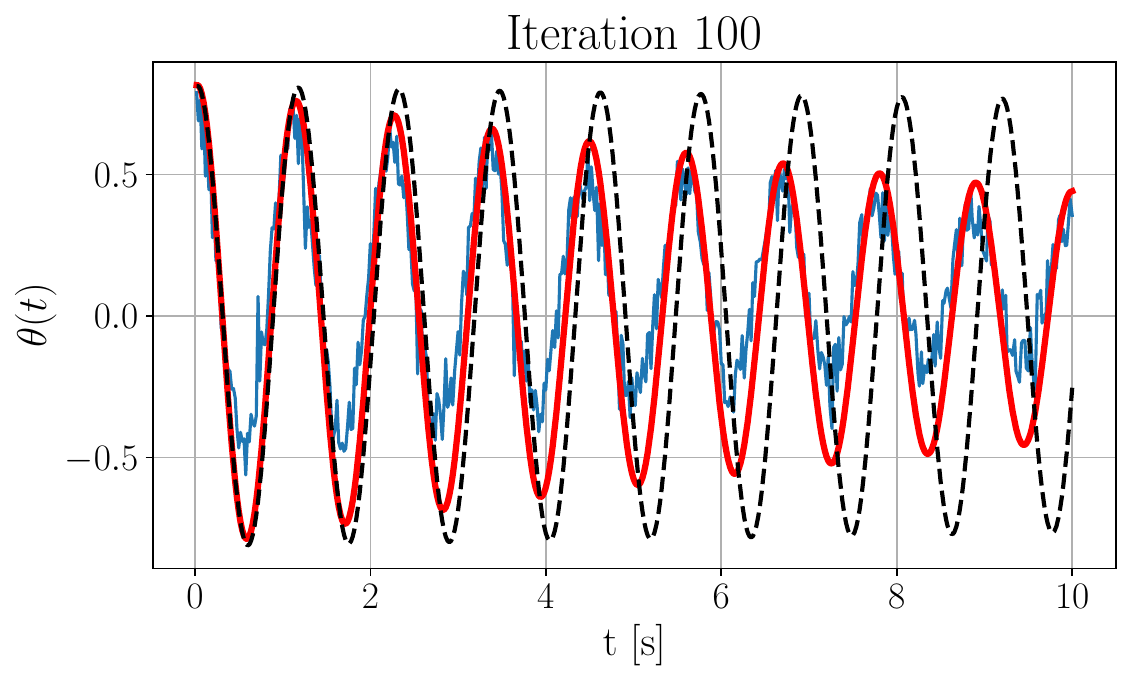}
         \caption{}
         \label{fig:ep_140_dyn}
     \end{subfigure}
     \hfill
     \begin{subfigure}[b]{0.42\linewidth}
         \centering
         \includegraphics[width=0.97\textwidth]{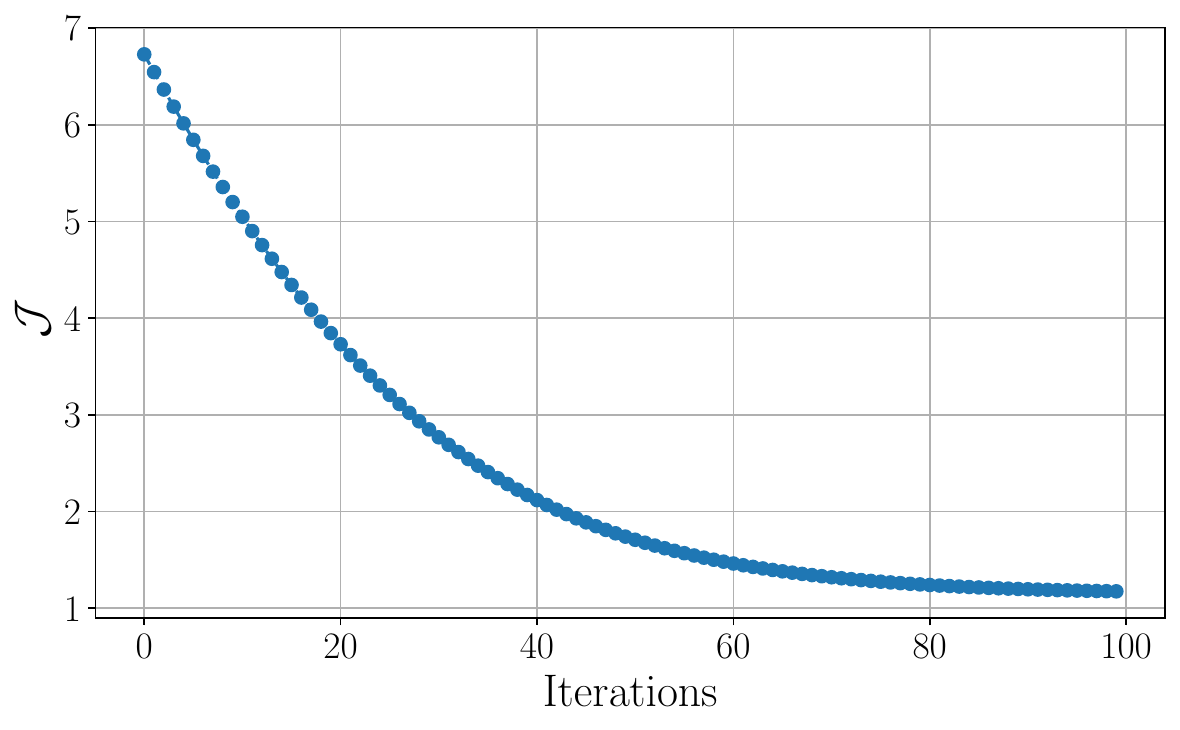}
         \caption{}
         \label{fig:ep_140_cost}
     \end{subfigure}
        \caption{Pendulum assimilation performances. The left-hand side shows the time-series of (a) the real system (blue), the initial guess (dashed black line), and the current assimilated prediction (solid red line). The right-hand side depicts the cost function history at the $i-$th iteration.}
        \label{fig:pendulum_ex_res}
\end{figure}

To take into account both noise and possibly unknown initial conditions, the code allows the definition of an ensemble average of the gradients considering a set of  $n_E$ virtual trajectories starting from different initial conditions and parametrized with the same choice of parameters $\bm{p} = \boldsymbol{w}_p$. Figure \ref{fig:pendulum_ex_res} shows the predictions of the digital twin versus the real system on the left, while the right panels show the evolution of the cost function while treating the data from each episode. The approach manages to assimilate the dynamical system reaching $J_p^{(i)}/J_p^{(0)}=$0.15 at the iteration $i$=100, despite the noise. Interestingly, for several episodes, the cost function occasionally increases; in these cases, the code stops the training and seeks another episode. More advanced strategies and possible combinations with more exploratory gradient-free methods are currently under development. Generally, however, the convergence performances are strictly linked with the smoothness of the cost function and may face challenges in non-smooth or complex optimization landscapes. In addition, the adjoint is very sensitive to initial conditions, as is the case with all gradient-based approaches tackling non-convex problems. Indeed, starting from a slightly different condition may compromise the trajectory evolution, possibly leading to inaccurate gradient estimates. In this exercise, we have shown a very naive way to mitigate this, by an ensemble approach over a set of possible initial states, but sometimes this may not be enough. 

\section{Conclusions and outlook}

This chapter offered hands-on experience with physics-constrained regression through three tutorial exercises. Though simplified, these examples highlighted both the potential and the core challenges of hybrid modeling. The differentiability of all presented models enabled the use of differential penalties and constraints, essential for tasks like data assimilation and modeling in boundary and initial value problems. However, the exercises also exposed key limitations: (1) RBF methods lead to dense, ill-conditioned systems, and ANN models depend on optimization algorithms that still lag behind traditional solvers; (2) effective training often requires careful architectural constraints to avoid vanishing gradients; and (3) results remain sensitive to initial conditions. These challenges mark active research frontiers, requiring engineers trained in both fluid mechanics \emph{and} machine learning.

%%%%%%%%%%%%%%%%%%%%%%%%%%
% USER ENTRY OFF
\bibliographystyle{apalike}

%\bibliographystyle{plainnat-doi}

% this is a nice bibliographics style
% USER ENTRY ON
\bibliography{chapter_3}
% USER ENTRY OFF
%\clearpage{\pagestyle{empty}\cleardoublepage}
% USER ENTRY ON
%\input{appendix}
% comment this input if you do not have an appendix
% USER ENTRY OFF
\end{document}